\documentclass[aps,twocolumn,floatfix,altaffilletter,superscriptaddress,preprintnumbers,tightenlines,showpacs,showkeys,notitlepage,nofootinbib]{revtex4-1}

\usepackage[T1]{fontenc}
\usepackage[colorlinks=true,citecolor=blue,linkcolor=blue]{hyperref}
\usepackage{braket,graphicx,physics,amsthm,amssymb,amsfonts,xcolor,booktabs,siunitx,caption,subcaption,multirow,tablefootnote}
\usepackage{mathtools}
\usepackage{slashed}
\usepackage[capitalise, english]{cleveref}
\usepackage[normalem]{ulem}
\usepackage{xspace}
\usepackage{mathrsfs}
\usepackage{soul}
\usepackage{subcaption}
\usepackage{fontawesome}
\newcolumntype{C}[1]{>{\centering\let\newline\\\arraybackslash\hspace{0pt}}m{#1}}

\usepackage{tikz}
\usetikzlibrary{arrows,shapes}
\usetikzlibrary{trees}
\usetikzlibrary{snakes}
\usetikzlibrary{matrix,arrows} 													
\usetikzlibrary{positioning}				
\usetikzlibrary{calc,through}				
\usetikzlibrary{decorations.pathreplacing} 
\usepackage[tikz]{bclogo} 					
\usepackage{pgffor}							
\usetikzlibrary{decorations.pathmorphing}	
\usetikzlibrary{decorations.markings}
\usetikzlibrary{intersections}		

\tikzset{
    vector/.style={decorate, decoration={snake}, draw},
    fermion/.style={postaction={decorate},
        decoration={markings,mark=at position .55 with {\arrow{>}}}},
    fermionbar/.style={draw, postaction={decorate},
        decoration={markings,mark=at position .55 with {\arrow{<}}}},
    fermionnoarrow/.style={},
    gluon/.style={decorate,
        decoration={coil,amplitude=4pt, segment length=5pt}},
    scalar/.style={dashed, postaction={decorate},
        decoration={markings,mark=at position .55 with {\arrow{>}}}},
    scalarbar/.style={dashed, postaction={decorate},
        decoration={markings,mark=at position .55 with {\arrow{<}}}},
    scalarnoarrow/.style={dashed,draw},
	vectorscalar/.style={loosely dotted,draw=black, postaction={decorate}},
}

\makeatletter
\providecommand*{\diff}%
	{\@ifnextchar^{\DIfF}{\DIfF^{}}}
\def\DIfF^#1{%
	\mathop{\mathrm{\mathstrut d}}%
		\nolimits^{#1}\gobblespace}
\def\gobblespace{%
	\futurelet\diffarg\opspace}
\def\opspace{%
	\let\DiffSpace\!%
	\ifx\diffarg(%
		\let\DiffSpace\relax
	\else
		\ifx\diffarg[%
			\let\DiffSpace\relax
		\else
  			\ifx\diffarg\{%
				\let\DiffSpace\relax
			\fi\fi\fi\DiffSpace}

\definecolor{cborange}{HTML}{e69f00}
\definecolor{cbgreen}{HTML}{009e73}
\definecolor{cbyellow}{HTML}{f1dd42}
\definecolor{cblblue}{HTML}{56b4e9}
\definecolor{cbblue}{HTML}{0072b2}
\definecolor{defgrey}{HTML}{9f9f9f}
\definecolor{defgreen}{HTML}{8eba42}

\newcommand{\muthreee}{Mu3e\xspace} 
 
\def\l@subsection#1#2{}
\def\l@subsubsection#1#2{}

\begin{document}


\title{Angling for Insights: Illuminating Light New Physics\\ at Mu3e through Angular Correlations}

\author{Simon Knapen}
\affiliation{Theory Group, Lawrence Berkeley National Laboratory, Berkeley, CA 94720, U.S.A.}
\affiliation{Berkeley Center for Theoretical Physics, University of California, Berkeley, CA 94720, U.S.A.}

\author{Kevin Langhoff}
\affiliation{Berkeley Center for Theoretical Physics, University of California, Berkeley, CA 94720, U.S.A.}
\affiliation{Theory Group, Lawrence Berkeley National Laboratory, Berkeley, CA 94720, U.S.A.}

\author{Toby Opferkuch}
\affiliation{Berkeley Center for Theoretical Physics, University of California, Berkeley, CA 94720, U.S.A.}
\affiliation{Theory Group, Lawrence Berkeley National Laboratory, Berkeley, CA 94720, U.S.A.}

\author{Diego Redigolo}
\affiliation{INFN, Sezione di Firenze Via G. Sansone 1, 50019 Sesto Fiorentino, Italy}

\begin{abstract}
We examine the capability of \muthreee to probe light new physics scenarios that produce a prompt electron-positron resonance and demonstrate how angular observables are instrumental in enhancing the experimental sensitivity. We systematically investigate the effect of \muthreee's expected sensitivity on the parameter space of the dark photon, as well as on axion-like particles and light scalars with couplings to muons and electrons. 
\end{abstract}


\maketitle


\section{Introduction}

The coming decade promises impressive progress in muon physics, with a host of new experiments that will be coming online or are already taking data. The muon $g-2$ experiment has reached an unprecedented level of precision on the muon magnetic dipole moment \cite{Muong-2:2023cdq} and it will soon be complemented by a suite of new probes of charged lepton flavor violation (CLFV). Experiments such as Mu2e \cite{Mu2e:2014fns}, COMET \cite{COMET:2018auw} and DeeMe \cite{Teshima:2019orf} focus on $\mu \to e$ conversions, while MEG II \cite{MEGII:2018kmf} and \muthreee~\cite{Mu3e:2020gyw,Hesketh:2022wgw} will be hunting for exotic muon decay channels as the hallmarks of beyond Standard Model (SM) physics. While this program has originally been envisioned as a probe of heavy new physics at mass scales as high as \qtyrange{e+4}{e+5}{\TeV}, see e.g.~\cite{Kuno:1999jp,Okada:1999zk,Calibbi:2017uvl,Davidson:2020hkf,Bolton:2022lrg}, it also provides a unique opportunity to test light new particles with extremely feeble couplings to the SM, see e.g.~\cite{Echenard:2014lma,Calibbi:2020jvd,Galon:2022xcl,Jho:2022snj,Hostert:2023gpk,Hill:2023dym}.

\begin{figure}
         \centering
         \includegraphics[scale = 1]{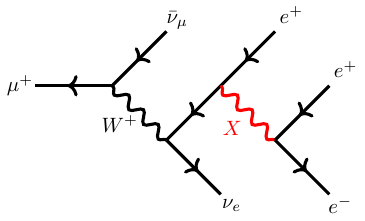}
         \caption{Example diagram of a light new particle $X$ produced in muon decays. Depending on the model, emissions from the muon and/or the neutrinos are also considered.}
         \label{fig:Flavor Conserving Diagram}
         
        \label{fig:Feynman Diagrams}
\end{figure}

In this work we zoom in on searches for light, weakly coupled particles at \muthreee that decay promptly to an $e^+e^-$ pair \cite{Echenard:2014lma,Perrevoort:2018ttp}. The \muthreee experiment~\cite{Mu3e:2020gyw,Hesketh:2022wgw} at the Paul Scherrer Institute (PSI) in Switzerland will study around \num{2.5E+15}{} muon decays during phase-I of operations and \num{5.5E+16}{} muon decays during phase-II with the primary goal of increasing the sensitivity to the CLFV decay $\mu^+ \rightarrow e^+e^+ e^-$. We focus our attention on the lepton flavor conserving decay $\mu^+ \to e^+ \bar \nu_\mu \nu_e X$, where the new particle $X$ can be radiated from any charged SM particle before decaying back to a dilepton pair ($X\to e^+ e^- $). This results in three tracks plus missing energy, as exemplified in Fig.~\ref{fig:Feynman Diagrams}. This channel was already considered in \cite{Echenard:2014lma,Perrevoort:2018ttp} with the main motivation of extending the reach on the parameter space of the minimal dark photon in the mass range between approximately \qtyrange{10}{100}{\MeV}. In this paper we expand on these previous analyses in two ways: \emph{i)} We broaden the physics case for \muthreee by considering different new physics models that can be probed in $\mu^+ \to e^+ \bar \nu_\mu \nu_e X$. \emph{ii)} We further study the impact of angular correlations in distinguishing signal from background, and show that a few simple variables can already deliver an $\mathcal{O}(1)$ improvement to the sensitivity without the need for modifications to the detector or data taking strategy. In the event of a discovery, angular correlations, including those obtained when accounting for muon polarization, can also be used to pin down the properties of the new particle. 

The main SM background originates from radiative muon decay ($\mu^+\to e^+ \bar \nu_\mu \nu_e\gamma^*$) where the off-shell SM photon splits into an electron-positron pair \mbox{($\gamma^*\to e^+e^-$)}. This process was computed at leading order in Ref.~\cite{Bardin:1972qq,Flores-Tlalpa:2015vga} and more recently at next-to-leading order in Ref.~\cite{Fael:2016yle,Pruna:2016spf,Banerjee:2020rww} and constitutes an irreducible background for all the signals considered here, besides also being the main background for the flagship CLFV \muthreee search: $\mu^+\rightarrow e^+e^+ e^-$. For the CLFV decay, the absence of missing energy allows one to impose very effective kinematic cuts, e.g.~by demanding that the three tracks reconstruct the invariant mass of the muon. Such a cut is much too strong for the models we consider here, and differentiating $\mu^+ \to e^+ \bar \nu_\mu \nu_e X$ from the background therefore requires a different strategy. In particular, a search for a resonance in the invariant mass distributions of the $e^+e^-$ pairs is already highly effective \cite{Echenard:2014lma,Perrevoort:2018ttp}, but nevertheless a large irreducible SM background remains. In this work we show that the sensitivity of a vanilla `bump hunt' analysis can be enhanced by an $\mathcal{O}(1)$ factor, by considering additional correlations in the angles and energies of the final-state charged leptons. 

We apply our analysis to light new physics scenarios where the new particle $X$ is either spin-0 or spin-1. For the purpose of our analysis, we can parametrize the interactions of $X$ with muons and electrons as 
\begin{align}
\label{eq:Lagrangians}
\mathscr{L}_{0}&\supset   X \left[\bar{\mu} (g^{\mu}_S+g^{\mu}_P\gamma_5) \mu+ \bar{e} (g^{e}_S+g^{e}_P\gamma_5) e\right]\,,\notag\\
\mathscr{L}_{1}&\supset    X^\alpha\left[\bar{\mu} \gamma_\alpha (g^{\mu}_V+g^{\mu}_A\gamma_5) \mu+  \bar{e} \gamma_\alpha (g^{e}_V+g^{e}_A\gamma_5) e\right]\,,
\end{align}
where for the scalar model, the couplings $g^{\ell}_{S}$($g^{\ell}_{P}$) parametrize the strength to parity even (odd) operator. Similarly, the $g^{\ell}_{V}$($g^{\ell}_{A}$) parametrize the strength of a spin-1 $X^\alpha$ to the vector (axial) current. We neglected extra possible interactions of $X$ with neutrinos, whose phenomenological relevance for \muthreee will be discussed in Appendix~\ref{app:additional_results}. We further assumed that interactions of the scalar $X$ with photons are suppressed relative to its interaction with the charged leptons. After obtaining results and physical insights with the general setup in Eqs.~\eqref{eq:Lagrangians}, we map those results onto motivated models, and compare with existing bounds. 

The remainder of this paper is organized as follows: In Sec.~\ref{sec:analysis-strategy} we briefly describe our simulation setup and analysis strategy. We then discuss in Sec.~\ref{sec:angles} the physical origin of the kinematical variables which allow us to improve the discrimination between signal and background. In Sec.~\ref{sec:results} we will discuss explicit models that fit in the generic parametrization of Eqs.~\eqref{eq:Lagrangians} and compare our forecasted reach at \muthreee with current constraints. We complement the discussion in the appendices: Additional results and models are presented in Appendix~\ref{app:additional_results}, followed by a discussion on the effects of the muon polarization in Appendix~ \ref{app:Polarization}. In Appendix~\ref{app:validation} we give further details on our simulation framework and validate our approximate detector modeling with public experimental results. 

\section{Analysis strategy}
\label{sec:analysis-strategy}

In the relevant regime of couplings for \muthreee, the decay width of the new resonance is much smaller than the detector resolution. The main observable is therefore a narrow resonance in the invariant mass distributions of the $e^+ e^-$-pairs, while the dominant SM background is a smooth falling distribution from the internal conversion process $\mu^+ \to \bar{\nu}_\mu e^+ \nu_e (\gamma^* \to e^+e^-)$. In this section we describe our variables and general strategy to enhance the sensitivity of a plain resonance search; all details on our event generation and our modeling of the detector response are deferred to Appendix~\ref{app:validation}.

\begin{figure}
    \begin{center}
        \includegraphics[width = 0.5\textwidth]{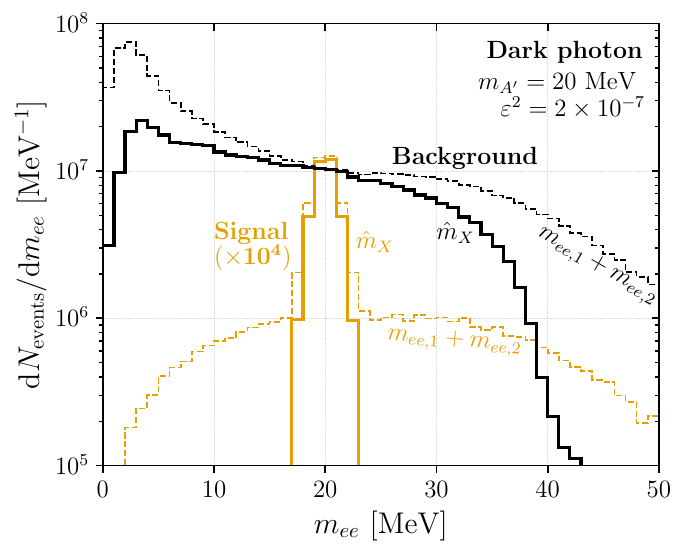}
    \end{center}
    \caption{Illustrative example of the signal ({\color{cborange}\bf orange lines}) and background ({\bf black lines}) distributions in the invariant mass plane of the electron-positron pair $m_{ee}$ after smearing and detector acceptance are applied, see \cref{app:validation} for details. In the notation of Eq.~\eqref{eq:Lagrangians}, we show the signal for the case of a vector spin-1 resonance with $g_V^\mu=g_V^e=e \epsilon$ (dark photon model) with mass $m_{X} = \SI{20}{\MeV}$ and coupling $\varepsilon^2 = 2\times 10^{-7}$ is multiplied by a factor $10^4$. Both the background and signal are shown for two cases: ($i$) the naive approach where one considers both the invariant mass combinations of the electron-positron pair ({\bf thin dashed} line labeled $m_{ee,1} + m_{ee,2}$) and ($ii$) our approach through the use of the observable $\hat{m}_X$, see \cref{eq:m_X}, which picks the combination closest to the mass hypothesis of the new resonance ({\bf thick solid} line labeled $\hat{m}_X$). }
    \label{fig:mee-plane-illustration}
\end{figure}

Since the final state contains two positrons, there is a combinatoric ambiguity in identifying the $e^+ e^-$ pair originating from the new resonance. We label the two possible invariant mass combinations of these pairs with $m_{ee,1}$ and $m_{ee,2}$. There are two main approaches to mitigating the combinatoric background. The first is to treat both combinations as individual events, effectively doubling the number of events in the $m_{ee}$ distribution. Once smearing and detector effects are included this can lead to a small number of spurious signal events where, depending on the value of $m_X$, the invariant mass of the electron with the second positron may also be close to $m_X$. This is shown in \cref{fig:mee-plane-illustration}, where the thin dashed lines show signal and background distributions utilizing this method. 

The second approach, which we will adopt in what follows, involves choosing the $e^+ e^-$ pair for which the invariant mass is closest to an arbitrarily chosen hypothesized signal mass $m_X$. In other words, we define the variable
\begin{equation} \label{eq:m_X}
\hat m_X \equiv \left\{ 
\begin{array}{ll} 
    m_{ee,1} & \text{for } |m_{ee,1}-m_X| < |m_{ee,2}-m_X|\,,\\
    m_{ee,2} & \text{for } |m_{ee,1}-m_X| > |m_{ee,2}-m_X|\,.
\end{array}\right.
\end{equation}
This choice automatically maximizes the signal efficiency, but artificially skews the background distribution by creating a feature centered around $m_{ee} = m_X$. This feature (solid black line in \cref{fig:mee-plane-illustration}) is however extremely broad compared to both the resonance width and the mass resolution of the detector. We therefore do not expect it to be a significant source of systematic uncertainties in a data-driven background estimation. 

We define $p_X$ as the four-vector of the $e^+ e^-$ pair with invariant mass $\hat{m}_X$ and label the positron assigned to the hypothesized resonance [through Eq.~\eqref{eq:m_X}] with $e^+_X$. The remaining, `solitary' positron is denoted by $e_s^+$. This means that whether a positron is assigned as `$e^+_X$' or `$e^+_s$' depends on the hypothesized value for $m_X$. With this convention in mind, we label the four-momentum of `solitary' positron $p_{e_s^+}$ and the four-momentum of the positron assigned to the resonance with $p_{e_X^+}$. The electron four-momentum is denoted by $p_{e^-}$. We further define the track energies normalized to the muon mass $m_\mu$ through
\begin{equation}
x_{e^-}\equiv \frac{2 E_{e^-}}{m_\mu}\,,\quad x_{e_X^+}\equiv \frac{2 E_{e_X^+}}{m_\mu}\,, \,\,\, \text{and}\,\,\, x_{e_s^+}\equiv \frac{2 E_{e_s^+}}{m_\mu}\,,
\end{equation}
which take values between 0 and 1.

In the next section, we will describe how angular correlations between these three tracks can help discriminating the signal from the background. Concretely, the variables we will need are
\begin{align}
\cos\theta_{e^+_s e^+_X}&\equiv \vec p_{e_s^+} \cdot \vec p_{ e^+_{X}} / (|\vec p_{e_s^+}| |\vec p_{e^+_{X}}|)\,,\\
\cos \theta_{ e^-e_s^+}&\equiv \vec p_{e_s^+} \cdot\vec p_{ e^-} / (|\vec p_{e_s^+}| |\vec p_{e^-}|)\,,
\end{align}
where the vector arrows label three-momenta. There is additional angular information if the polarization of the muon beam is taken into account; however, it turns out that these additional angles yield less distinguishing power. They are however more sensitive to the differences between the models presented in the previous section, and are therefore well suited to distinguish models from one another in the event of a positive discovery. We defer this discussion to Appendix~\ref{app:Polarization}.

A rigorous experimental analysis would presumably construct a global likelihood for the full kinematics of both signal and background. Here we opt for a simplified approach, which is computationally faster and has the advantage that the effect of the angular correlations can be isolated easily. Concretely, we first place a cut on the $\hat m_{X}$ distribution in a symmetric window around our hypothesis for $m_X$. The optimal size of this window is determined by maximizing $S/\sqrt{B}$, with $S$ and $B$ the number of signal and background events for a given window size. We subsequently bin the events in this window in two distinguishing variables from the set of $\{ \cos\theta_{e^-e_s^+},\,\cos\theta_{e^-e_X^+},\, x_{e_s^+}\}$. We must limit ourselves to binning in two dimensions because of size limitations of our background Monte Carlo sample, while in Sec.~\ref{sec:angles} we discuss why these variables are the most effective in separating the signal from the background.

To estimate the additional sensitivity provided by any chosen pair of distinguishing variables we construct a binned Poisson likelihood over both variables
\begin{align}\label{eq:likelihood}
    \mathcal{L}(S_i,B_i,n_i) &= \prod_{i}\frac{ \left( S_i + B_i \right)^{n_i}}{n_{i} !} e^{-\left(S_i + B_i \right)}\,.
\end{align}
Here $S_i$ ($B_i$) are the bin-wise signal (background) expectation values, $n_i$ are the number of observed events in the $i$-th bin.  The $n_i$ are drawn from the background-only distribution ($S_i=0$). As our test statistic, we then define the log-likelihood ratio using Eq.~\eqref{eq:likelihood}
\begin{align}
\lambda &\equiv -2\log\left[\frac{\mathcal{L}(S_i,B_i,n_i)}{\mathcal{L}(S_i=0,B_i,n_i)}\right]\approx \sum_i S_i^2/B_i\,. \label{eq:lambda_approx}
\end{align}
In the high statistics limit, the Poisson likelihood in Eq.~\eqref{eq:likelihood} asymptotes to a Gaussian and the likelihood ratio can be approximated by the second equality in Eq.~\eqref{eq:lambda_approx}, where we took $n_i\approx B_i$. To extract the 95\% expected exclusion limit (one-sided), we solve for $\lambda=2.71$. We performed this procedure both with the Poisson and Gaussian likelihoods, and established that the latter is always an excellent approximation. Additional details and validation figures can be found in \cref{app:validation}.

\section{Angular correlations}
\label{sec:angles}

\begin{figure*}
    \centering
    \begin{subfigure}{0.24\textwidth}
        \includegraphics[width=\linewidth]{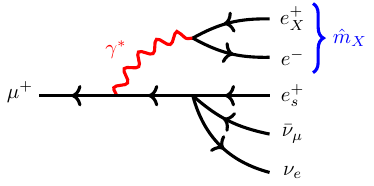}
        \caption{Diagram with $\mathcal{B}_\mu$}
        \label{fig:B_mu}
    \end{subfigure}
    \hspace{0.0\textwidth}
        \begin{subfigure}{0.24\textwidth}
        \includegraphics[width=\linewidth]{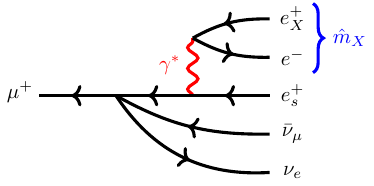}
        \caption{Diagram with $\mathcal{B}_e$}
        \label{fig:B_e}
    \end{subfigure}
    \begin{subfigure}{0.24\textwidth}
        \includegraphics[width=\linewidth]{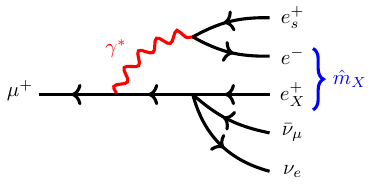}
        \caption{Diagram with $\mathcal{B}_\mu'$}
        \label{fig:B_mu_Prime}
    \end{subfigure}
    \begin{subfigure}{0.24\textwidth}
        \includegraphics[width=\linewidth]{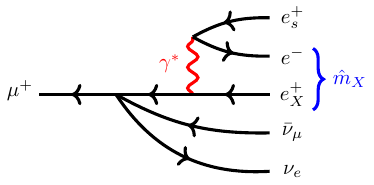}
        \caption{Diagram with $\mathcal{B}_e'$}
        \label{fig:B_e_Prime}
    \end{subfigure}
    \caption{Feynman diagrams for the SM background in Eq.~\eqref{eq:background_poles}. The left two diagrams identify the electron-positron pair with invariant mass closest to $\hat{m}_X$ as those that came from the virtual photon. The right two diagrams identify the other combination. It is these second two diagrams which are exclusive to the background leading to an altered distribution in $\cos \theta_{e^-e_s^+}$ compared to the signal.}
    \label{fig:background_diagrams}
\end{figure*}

Because of the difference in matrix elements, we expect that the full angular distribution  may provide additional signal versus background discrimination. Making use of all this information is however non-trivial, as the statistics of any realistic Monte Carlo sample will tend to be smaller than the number of actual background events \muthreee will record. This means that it is difficult to build a reliable, high-dimensional likelihood function without a data-driven background estimation. 

We address this difficulty in two ways: Firstly, we implement a custom set of cuts in Madgraph, which allows us to generate a separate, high-statistics background sample tailored to each signal model point. This method is described and validated in~\cref{app:validation}. Secondly, we analyze the amplitudes of both signal and background to identify a small set of the most promising variables with which to perform a likelihood analysis. In this section we will describe the physics behind these variables.

\subsection{Model-independent Variable}

Before we discuss any dynamics related to a specific model, it is valuable to consider first the general features of both signal and background amplitudes. Note that for the sake of clarity we will set the electron mass to zero in what follows, though $m_e$ was set to its physical value for all our numerical results. The general form of the amplitude for the signal process, irrespective of the specific model considered, will be 
\begin{align}\label{eq:signal_poles}
    \mathcal{M}_{\rm sig} = \frac{\mathcal{S}_\mu}{(p_\mu - p_{X})^2 - m_\mu^2} + \frac{\mathcal{S}_e}{(p_X + p_{e^+_s})^2}\,. 
\end{align}
Here the first and second terms correspond to the case where $X$ is radiated off the final state $e_s^+$ and initial state $\mu^+$, respectively, while $\mathcal{S}_{e/\mu}$ are the numerators of the amplitude which are model dependent but do not have additional poles.

This must be compared to the background which also has a propagator from the virtual photon. Concretely, there are four diagrams, shown in Fig.~\ref{fig:background_diagrams}, which correspond to an amplitude of the form
\begin{align}\label{eq:background_poles}
\mathcal{M_{\rm bkg}} &=\frac{1}{(p_\mu - p_X )^2 - m_\mu^2}\left[\frac{\mathcal{B}_\mu}{p_{e^-}\cdot p_{e^+_X}} +\frac{\mathcal{B}_\mu'}{p_{e^-}\cdot p_{e^+_s}} \right] \notag \\
&\quad+\frac{1}{(p_X + p_{e^+_s})^2}\left[\frac{\mathcal{B}_e}{p_{e^-}\cdot p_{e^+_X}} +\frac{\mathcal{B}_e'}{p_{e^-}\cdot p_{e^+_s}} \right]\,.
\end{align}
The $\mathcal{B}_\mu$, $\mathcal{B}'_\mu$, $\mathcal{B}_e$ and $\mathcal{B}'_e$ terms correspond respectively to the diagrams \ref{fig:B_mu}, \ref{fig:B_mu_Prime}, \ref{fig:B_e} and \ref{fig:B_e_Prime} in Fig.~\ref{fig:background_diagrams}. If we investigate the pole structure in Eq.~\eqref{eq:background_poles} we find
\begin{align}
    p_{e^-}\cdot p_{e_X^+} &= m_X^2/2\,,\label{eq:massinv}\\
    p_{e^-}\cdot p_{e_s^+} &= E_{e^-}E_{e_s^+}(1-\cos \theta_{e^- e_s^+})\,,
\end{align}
which implies that the terms which scale as $(p_{e^-}\cdot p_{e_s^+})^{-1}$ will dominate the amplitude when $e^-$ and $e_s^+$ become collinear. This feature is most pronounced for large $m_X$ where the other terms are suppressed by virtue of Eq.~\eqref{eq:massinv}.

\begin{figure*}
\centering
   \begin{subfigure}{0.328\textwidth}
        \includegraphics[width=1.0\linewidth,trim={0.3cm 0 0.2cm 0}, clip]{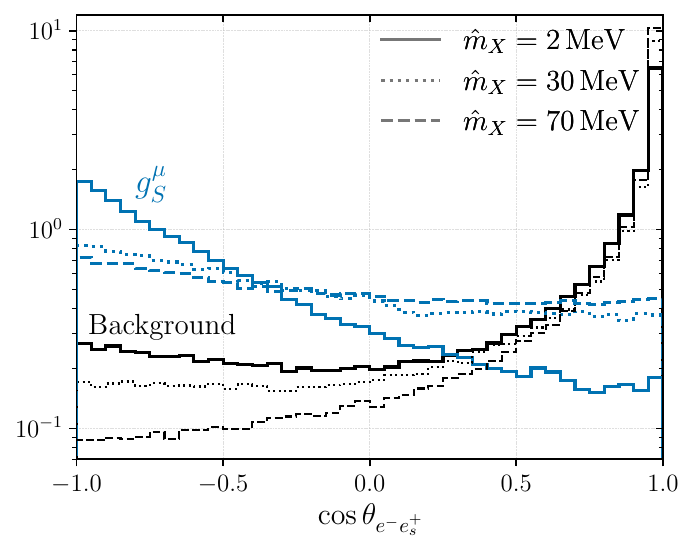}
        \caption{}
         \label{fig:cos_e_es}
    \end{subfigure}
    \hspace{-0.06cm}
    \begin{subfigure}{0.328\textwidth}
        \includegraphics[width=1.0\linewidth,trim={0.3cm 0 0.2cm 0}, clip]{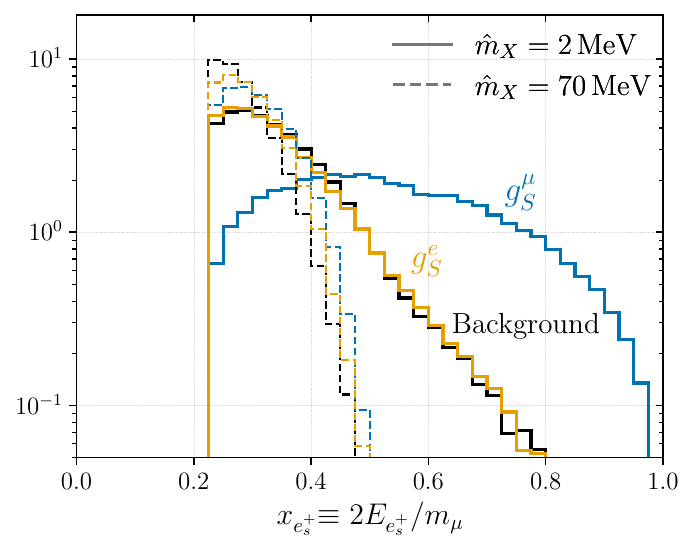}
        \caption{}
        \label{fig:xe}
    \end{subfigure}
     \hspace{-0.06cm}
    \begin{subfigure}{0.328\textwidth}
        \includegraphics[width=1.0\linewidth,trim={0.3cm 0 0.2cm 0.2cm}, clip]{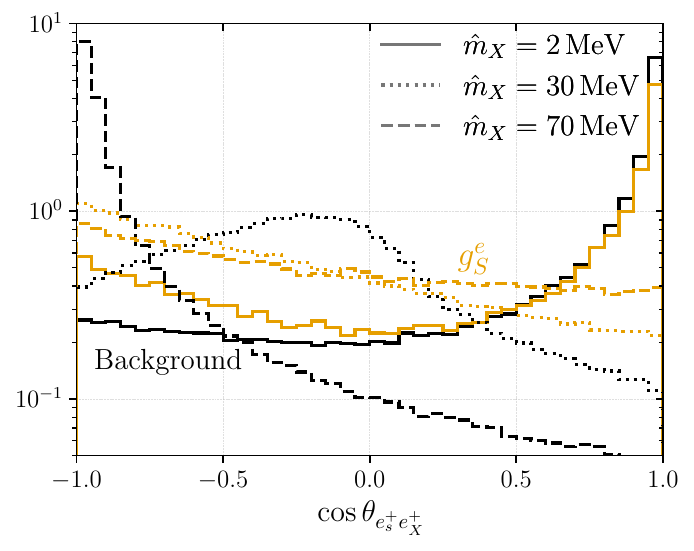}
        \caption{}
         \label{fig:cos_es_ex}
    \end{subfigure}
\caption{
Normalized distributions for the background ({\bf black lines}) and signal for a scalar coupled predominantly to muons $g_S^\mu \gg g_S^e\neq 0$ ({\color{cbblue}\bf blue lines}) and for a scalar coupled to electrons $g_S^e\neq 0$) ({\color{cborange}\bf orange lines}) for the kinematic variables included in our analysis. For the background distributions we require $\hat m_X$ to be within a window of $\pm \SI{2.5}{\MeV}$ around the respective value quoted in the legend. All distributions are normalized to one.}
\end{figure*}

The pole at \mbox{$\cos \theta_{e^- e_s^+} = 1$} is due to the off-shell SM photon in \cref{fig:B_mu_Prime,fig:B_e_Prime} which is absent for all signal models. As a result $\cos \theta_{e^- e_s^+}$ is a powerful model-independent variable to help separate signal and background. This is shown in Fig.~\ref{fig:cos_e_es}, which shows a clear peak for the background (black lines) across all values of $\hat{m}_X$. For illustration we also show a particular signal model (blue lines corresponding to a scalar coupled predominantly to muons $g_S^\mu \gg g_S^e\neq 0$) which is anti-correlated with the background in part due to the absence of the same pole.

\subsection{Model-dependent Variables}
To discuss the angular variables that are model-dependent we consider a CP-even scalar coupled predominantly to either muons or electrons (through $g^\mu_S$ or $g^e_S$ in Eq.~\eqref{eq:Lagrangians}). However, the observations discussed here depend only on whether the muon or the electron coupling dominates and will therefore also apply to the case of the CP-odd scalar.

The squared amplitudes for a scalar coupled dominantly to electrons or muons are respectively
 \begin{align} \label{eq:M2-electron-coupling}
    \frac{1}{2}\sum_\text{spins}|\mathcal{M}_e|^2 &= 16 G_F^2 (g^{e}_{S})^2\frac{|\mathcal{M}_{e,1}|^2|\mathcal{M}_{e,2}|^2}{(p_X+p_{e^+_s} )^{4}}\ ,\\
    \frac{1}{2}\sum_\text{spins}    |\mathcal{M_\mu}|^2 &= 16 G_F^2 (g^{\mu}_{S})^2\frac{|\mathcal{M}_{\mu,1}|^2|\mathcal{M}_{\mu,2}|^2}{\left[(p_\mu - p_X)^2-m_\mu^2\right]^2}\ , 
    \label{eq:M2-muon-coupling}
\end{align}
with
\begin{align}
  |\mathcal{M}_{e,1}|^2 &= 2 E_{\nu_e} m_\mu\ , \notag\\
  |\mathcal{M}_{e,2}|^2 &= 4 (p_X \cdot p_{e^+_s})(p_X\cdot p_{\nu_\mu})- 2 m_X^2 (p_{e^+_s} \cdot p_{\nu_\mu})\ ,\notag\\
  |\mathcal{M}_{\mu,1}|^2 &= 4m_\mu(E_X-2m_\mu)(p_X\cdot p_{\nu_e})\notag\\
  &\qquad+8 m_\mu E_{\nu_e}  (m_\mu^2-\frac{1}{4}m_X^2) \ , \notag\\
  |\mathcal{M}_{\mu,2}|^2 &= 2 p_{e_s^+} \cdot p_{\bar\nu_\mu}\ .   
\end{align}
In Eq.~\eqref{eq:M2-electron-coupling} and Eq.~\eqref{eq:M2-muon-coupling} we recognize the pole structure we anticipated in Eq.~\eqref{eq:signal_poles}. Concretely, we can expand the denominators as
\begin{align}
    &(p_X+p_{e^+_s} )^2=2E_X E_{e^+_s}(1-\beta_X \cos \theta_{e_s^+X})+m_X^2\ , \label{eq:gee_denom}\\
    &(p_\mu - p_X)^2-m_\mu^2=m_X^2-2m_\mu E_X\ , \label{eq:gmumu_denom}
\end{align}
where $\beta_X$ is the velocity of $X$. Inserting Eq.~\eqref{eq:gee_denom} into Eq.~\eqref{eq:M2-electron-coupling} shows that at low $m_X$, the amplitude for the scalar coupled to electrons prefers $E_{e^+_s}$ to be as small as possible. The same pattern can be found in the background amplitude, looking at the second line in Eq.~\eqref{eq:background_poles}. This behavior is not present if the scalar $X$ couples predominantly to muons, as can be seen by plugging Eq.~\eqref{eq:gmumu_denom} in Eq.~\eqref{eq:M2-muon-coupling}. We therefore expect $E_{e^+_s}$ to be larger in the muon coupled case. This is shown in Fig.~\ref{fig:xe}, where we used the dimensionless energy variable $x_{e^+_s}$ defined in the previous section. In the low $m_X$ regime we therefore expect $x_{e^+_s}$ to be a useful variable for models which predominantly couple to muons.

With the variable above we have not yet fully exhausted all the angular information in the event. Using the angle $\cos \theta_{e_s^+e_X^+}$ additional discrimination between background and signal can be achieved. The $\cos \theta_{e_s^+e_X^+}$ distributions for signal and background are shown in Fig.~\ref{fig:cos_es_ex}, for different choices of $m_X$. For the background, we see that both positrons shift from being colinear for low $m_X$, to back-to-back configurations at higher masses. The former is easily understood in terms of the soft-colinear emission preferred by Eq.~\eqref{eq:background_poles}. For high $m_X$, the pieces scaling as $(m_X)^{-2}$ are suppressed but the remaining terms still prefer the $e^-$ and $e^+_s$ to be colinear. Momentum conservation then forces $e^+_X$ to be back-to-back with $e^+_s$ and $e^-$, as both neutrinos are very soft in this regime. A priori, one would therefore expect from Fig.~\ref{fig:cos_es_ex} that $\cos \theta_{e_s^+e_X^+}$ is the most powerful either at low or at high $m_X$. This is not the case however, as in both those limits this angle is strongly correlated with $\cos \theta_{e^- e_s^+}$ and it therefore adds little to the significance. It however does contribute at intermediate $m_X$. 

\begin{figure}
   \centering
   \includegraphics[width=\linewidth]{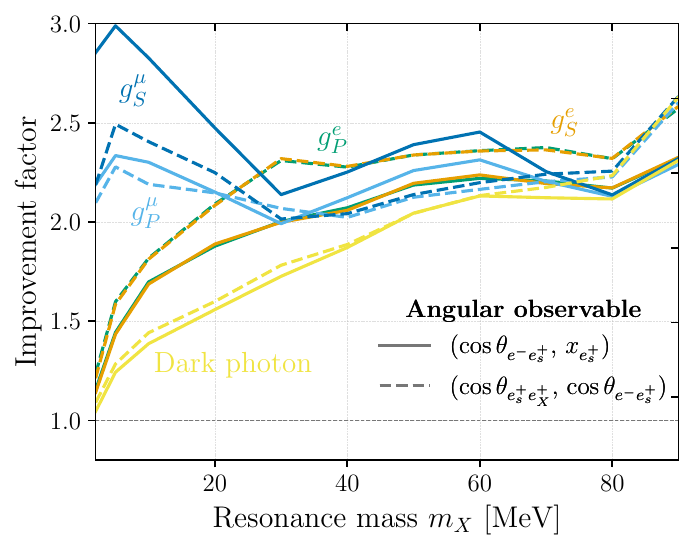}
   \caption{Improvement of the reach shown as a ratio of the rates (couplings squared) from including the two combinations of kinematic observables. The dark photon model ({\color{cbyellow}\textbf{yellow lines}}) used here as a benchmark refers to $g_V^\mu=g_V^e=e \epsilon$. The scalar ({\color{cbblue}\textbf{blue lines}}) or pseudoscalar ({\color{cblblue}\textbf{light-blue lines}}) coupled predominantly to muons refers to $g_{S,P}^\mu \gg g_{S,P}^e\neq 0$. While lastly the scalar ({\color{cborange}\textbf{orange lines}}) or pseudoscalar ({\color{cbgreen}\textbf{green lines}}) coupled only to electrons refers to $g_{S,P}^e\neq 0$, all of the above in the notation of \cref{eq:Lagrangians}.} 
   \label{fig:Improvement_Factor}
\end{figure}

\subsection{Summary}

To assess the enhancement in reach achievable by incorporating our new kinematic variables, we supplement the bump hunt strategy with a 2D likelihood in two pairs of variables: ($\cos \theta_{e^- e_s^+}$, $\cos \theta_{e_X^+ e_s^+}$) and ($\cos \theta_{e^- e_s^+}$, $x_{e_s^+}$). Subsequently, we normalize this outcome against the sensitivity obtained using a conventional bump-hunt method, resulting in the relative improvement in reach, as depicted in Fig.~\ref{fig:Improvement_Factor}. Here colors refer to the various signal models while the dashed versus solid lines refer to the pair of kinematical variables chosen. 

We observe an approximate twofold or better increase in reach across all models and mass selections, except in the case of the dark photon model at low $m_X$. In this low mass range, the dark photon signal closely resembles an off-shell SM photon exchange, leading to diminished sensitivity. Conversely, when increasing the signal resonance's mass, distinctions in angular correlations between the signal and background become more pronounced. In such scenarios, the variable $\cos \theta_{e^- e_s^+}$ predominantly contributes to sensitivity. Additionally, for (pseudo)scalar couplings to electrons, we find that the inclusion of $\cos \theta_{e_s^+ e_X^+}$ yields the most favorable results, whereas incorporating $x_{e^+_s}$ proves advantageous for (pseudo)scalar couplings to muons, as anticipated in the preceding section.

\section{Results} 
\label{sec:results}
In this section we describe a few benchmark models whose unconstrained parameter space can be explored through the decay $\mu^+ \rightarrow e^+ \bar{\nu}_{\mu} \nu_e (X \rightarrow e^+e^-)$. We discuss the dark photon in Sec.~\ref{sec:darkphoton}, light scalars and axion-like particles in Sec.~\ref{sec:lightscalar}. We comment on other models in Appendix~\ref{app:additional_results} where the \muthreee reach is likely not competitive with existing experiments.
\subsection{Dark photon}
\label{sec:darkphoton}
Dark photons are present whenever the standard model is embedded into a theory which contains an extra $U(1)$ gauge symmetry~\cite{Fayet:1980ad,Fayet:1980ad,Okun:1982xi,Georgi:1983sy,Holdom:1985ag,Essig:2013lka,Proceedings:2012ulb,Jaeckel:2010ni}. They feature in a variety of phenomenological models, such as those where it acts as a mediator particle to a dark sector~\cite{Pospelov:2008zw} as well as attempts to explain the the anomalous magnetic moment of the muon \cite{Fayet:2007ua}. 

At low energies, the interaction between the dark photon and the SM comes through the kinetic mixing with the SM photon 
\begin{align}
    \mathcal{L}\supset-\frac{1}{4} F_{ \mu \nu} F^{\mu \nu}-\frac{1}{4} F^{\prime}_{ \mu \nu} F^{\prime\, \mu \nu}-\frac{\varepsilon}{2} F_{ \mu \nu} F^{\prime\,\mu \nu}\,,
\end{align}
where the mixing parameter $\varepsilon$ can be naturally taken to be arbitrarily small. Upon diagonalization of these kinetic terms the dark photon interaction with SM fermions ($f$) are generated and the relevant Lagrangian for this paper is the following:
\begin{align}\label{eq: Dark Photon Lagrangian}
    \mathcal{L} = -\sum_{f}  \varepsilon q_f \bar{f} \gamma^\mu f A'_\mu+ m_{A^\prime}^2 A'_\mu {A'}^\mu \,,
\end{align}
with $q_f$ the electric charge of $f$. In the notation of Eq.~\eqref{eq:Lagrangians} the dark photon has $g_V^\mu=g_V^e=e\varepsilon$ and $g_A^\mu=g_A^e=0$ and its mass $m_{A'}$ can be considered as independent from its coupling to the SM. 

If the dark photon is lighter than the other potential states in the dark sector, it will decay back to SM particles through the interaction in Eq.~\eqref{eq: Dark Photon Lagrangian}. For $2m_e < m_{A'}< m_\mu$, its proper decay length is given by
\begin{equation}
c\tau_{A'} \approx 0.5\, \text{mm} \left(\frac{10^{-4}}{\epsilon}\right)^2  \left(\frac{\SI{10}{\MeV}}{m_{A'}}\right).
\end{equation}
As we will see, this means that the dark photon decay always occurs promptly compared the vertex resolution ($\sim \SI{3}{\mm}$) for the range of $m_{A'}$ and $\epsilon$ which is accessible to \muthreee but not constrained by other experiments such as NA64~\cite{Banerjee:2019hmi}. 

\begin{figure}
    \centering
    \includegraphics[width=1\linewidth]{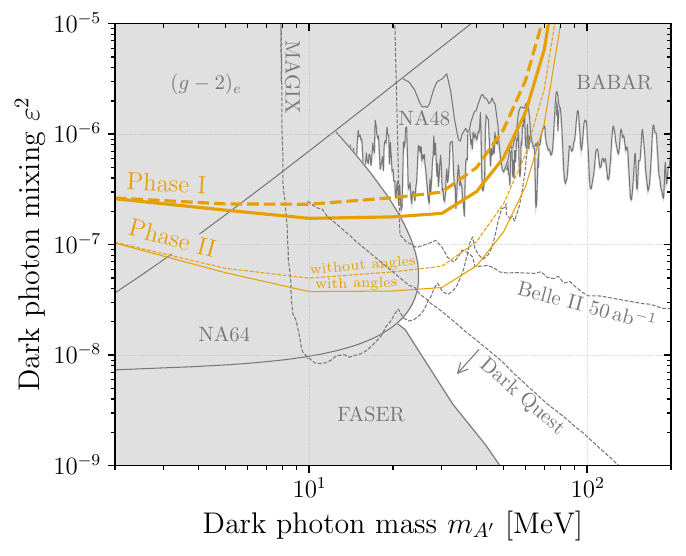}
    \caption{Projections for the 95\% upper limit on $\varepsilon^2$ ({\color{cborange}\bf orange lines}) for the dark photon model at \muthreee phase I $(2.5\times 10^{15} \mu^+)$ and phase II $(5.5\times 10^{16} \mu^+)$. The {\bf solid} and {\bf dashed} curves indicate our projected sensitivity with and without the inclusion of angular variables described in Sec.~\ref{sec:angles}. {\color{defgrey}\bf Gray shaded} areas are the existing exclusion limits from FASER \cite{Petersen:2023hgm}, NA64 \cite{Banerjee:2019hmi}, the electron magnetic dipole moment \cite{Bodas:2021fsy}, NA48 \cite{Batley:2015lha} and BABAR \cite{Lees:2014xha}. {\color{defgrey}\bf Thin dashed gray} lines indicate the expected sensitivity of Belle~II with the full luminosity of $50\text{ ab}^{-1}$~\cite{Belle-II:2018jsg}, the MAGIX~\cite{Doria:2018sfx} proposal at MESA and the DarkQuest proposal at Fermilab~\cite{Berlin:2018pwi}. The existing exclusions in this figure were generated with the DarkCast package \cite{Ilten:2018crw}.}
    \label{fig:Dark_Photon_Constriants}
\end{figure}

Our results for the dark photon model are shown in Fig.~\ref{fig:Dark_Photon_Constriants}, with and without the inclusion of the angular variables from the previous section. Without including the angular correlations, our results are in excellent agreement with prior work \cite{Perrevoort:2018ttp,Perrevoort:2018okj} for $m_{A'}\gtrsim \SI{30}{\MeV}$. For  $m_{A'}\lesssim 30$ MeV we find slightly weaker sensitivity, which is due to our Monte Carlo overestimating the background in this region (See Fig.~\ref{fig:bkg-validation} in Appendix~\ref{app:validation}). Such low masses are however already disfavored by limits from NA64. 

By including the angular information, we find the sensitivity to the rate, or equivalently to $\varepsilon^2$, can be improved by roughly a factor of two over the remaining mass range of interest. The enhanced reach means that phase I of \muthreee will be well positioned to cover the gap at low mass between NA64 and NA48. This gap has recently received a lot of attention because of the claimed light new physics interpretation of the ATOMKI anomaly~\cite{Krasznahorkay:2015iga} and for this reason the small mass region around 17 MeV will probably be tested before \muthreee phase I by dedicated experiments such as PADME~\cite{Nardi:2018cxi} and the special configuration of MEG~II~\cite{megIIatomki}. However, both these experimental configurations are expected to quickly lose sensitivity when the dark photon mass is changed, therefore, leaving a lot of unexplored parameter space to be probed by \muthreee already in phase I.

Phase II of \muthreee will carve out significantly more parameter space competing with other complementary probes such as Belle~II~\cite{Belle-II:2018jsg}, the proposed MAGIX experiment at the MESA accelerator in Mainz~\cite{Doria:2018sfx} and the DarkQuest proposal at Fermilab~\cite{Berlin:2018pwi}, as well as eventually LHCb~\cite{Ilten:2015hya}.

\subsection{Light scalars and axion-like particles}
\label{sec:lightscalar}
Light scalars and pseudoscalars in the \qtyrange{1}{100}{\MeV} mass range are ubiquitous in many extensions of the SM, ranging from dark sector models~\cite{Pospelov:2008zw,Batell:2017kty,Ballett:2018ynz}, heavy QCD axion models~\cite{DIMOPOULOS1979435,Holdom:1982ex,Flynn:1987rs,Rubakov:1997vp,Choi:1998ep} to generic models containing pseudo-Nambu-Goldstone bosons associated to spontaneously broken approximate global symmetries~\cite{Kilic:2009mi,Ferretti:2013kya,Belyaev:2015hgo,Bellazzini:2017neg,Ferretti:2016upr,Jeong:2018ucz}. Independently of the UV origin of the light scalar and its CP nature, we can ask if it is natural for it to be lighter than the muon mass and at the same time interacting with muons and/or electrons with a coupling strength large enough to be testable at \muthreee. Assuming that the couplings of the light scalar are generated by some UV physics at a scale $\Lambda_{\text{UV}}$, generically the expected one-loop correction to the light scalar mass induced by these couplings can be estimated as  
\begin{equation}
\delta m_X\sim \frac{g_{S,A}^{\mu,e}}{4\pi} \Lambda_{\text{UV}}\,.
\end{equation}
Requiring the above correction to be smaller or of the same order of the light scalar mass (i.e. $\delta m_X\lesssim m_X)$ we get a rough prediction for the scale at which new particles are expected to be present
\begin{equation}
\Lambda_{\text{UV}}\lesssim \SI{625}{\GeV}\left(\frac{\num{4e-4}}{g_{S,A}^{\mu,e}}\right)\left(\frac{m_X}{\SI{20}{\MeV}}\right)\,,\label{eq:naturalness}
\end{equation}
which shows that seeing light scalars within the reach of \muthreee is compatible with not having found heavy electroweak charged states at around or above electroweak scale.

For $2m_e<m_X<m_\mu-m_e$, the available visible decay channels are into di-lepton and di-photon pairs. The former decay width is directly controlled by the couplings introduced in Eq.~\eqref{eq:Lagrangians}. For $m_X\gg m_e$ the decay width of $X$ is given by
\begin{equation}
\Gamma_{X\to e^+ e^-} \approx \frac{g_e^2}{8\pi} m_X\ , 
\end{equation}
where $g_e\equiv \sqrt{(g_{S}^e)^2+(g_{A}^e)^2}$. The scalar decay width into diphotons is unavoidably loop-generated anytime the scalar couples to muon, resulting in the ratio of partial widths
\begin{equation}
\frac{\Gamma_{X\to \gamma\gamma}}{\Gamma_{X\to e^+ e^-} } \approx \num{2e-3}\left(\frac{m_X}{\SI{10}{\MeV}}\right)^2\left( \frac{g_\mu/g_e}{10^3}\right)^2 \,.
\end{equation}
This shows that the proper decay length of $X$ is dominated by the di-electron channel, even for a large hierarchy between $g_\mu$ and $g_e$. (This assumes that the diphoton width does not get significant contributions from UV physics.) Under these circumstances we can estimate the decay width of the scalar as 
\begin{equation} \label{eq:ctau_scalar}
c\tau_X \approx \SI{5}{\mm}\left(\frac{\num{e-5}}{g_e}\right)^2\left(\frac{\SI{10}{\MeV}}{m_X}\right)\,.
\end{equation}
\begin{figure*}
    \centering
    \includegraphics[width=0.495\linewidth]{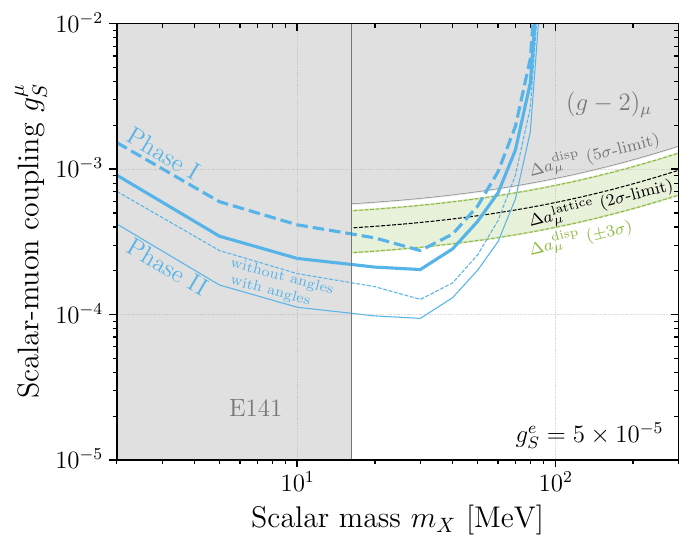}\hfill
    \includegraphics[width=0.495\linewidth]{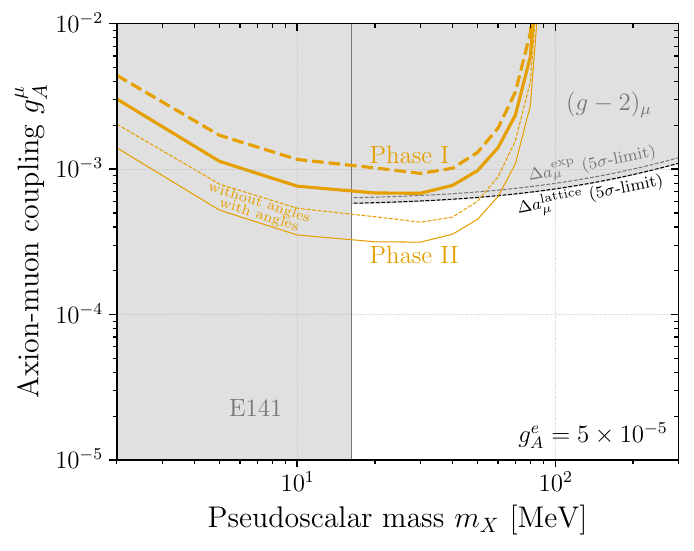}
    \caption{Parameter space of a  CP-even scalar ({\bf left}) and a CP-odd scalar ({\bf right}) with hierarchy between the muon and electron couplings smaller than the SM mass hierarchy: $g_{S,A}^e/g^\mu_{S,A}>m_e/m_\mu$. Projections for the 95\% upper limit on $g_S^\mu$ and  $g_P^\mu$ are shown in {\color{cbblue}\bf blue} and {\color{cborange}\bf orange}, respectively. The {\bf thick} lines correspond to phase I of \muthreee with $\num{2.5E+15}\mu^+$ and the {\bf thin} lines to the phase II with $\num{5.5E+16}\mu^+$. The {\bf solid} and {\bf dashed} lines show the reach with and without the use of angular correlations. The {\color{defgrey}{\bf gray shaded}} areas show constraints from Orsay~\cite{Riordan:1987aw} and the $(g-2)_\mu$ measurement. In the scalar case the best-fit region for $\Delta a_\mu^{\text{disp}}$ defined in Eq.~\eqref{eq:disp} is shown as a {\color{defgreen}{\bf green}} shaded region while the limit from $\Delta a_\mu^{\text{lattice}}$ defined in Eq.~\eqref{eq:lattice} is indicated as a {\bf thin-dashed black} line.}
    \label{fig:gaee-constraints}
\end{figure*}
A ``prompt'' search at \muthreee requires the $X$ decay to occur within $\SI{3}{\mm}$ of the corresponding $\mu$ decay, which means that this strategy is limited to $g_{e}\gtrsim 10^{-5}$ before events begin to be lost due to vertexing.
Within this range of couplings it makes sense to ask where a prompt search at \muthreee can have the largest impact. In general, the physics capabilities of \muthreee are most unique when the electron couplings of $X$ are substantially smaller than the muon couplings. This is because stringent bounds from electron beam dumps~\cite{Davier:1989wz,Bjorken:1988as} probing the same region of parameter space are relaxed.   

In Fig.~\ref{fig:gaee-constraints} we show a benchmark scenario where the electron couplings are suppressed with respect to the muon, but not to the extent that they scale as the ratio of the electron mass over the muon mass  $(g_{S,A}^e/g^\mu_{S,A}>m_e/m_\mu)$. In this scenario the muon coupling dominates the signal branching ratio while the electron coupling is sufficiently small to not be constrained (see Fig.~\ref{fig:gaee-constraints_mass_hierarchy}) but sufficiently large to produce prompt decays at \muthreee. One should keep in mind that generating the scalar interactions in Eq.~\eqref{eq:Lagrangians} with arbitrary hierarchies needs some amount of alignment in the ultraviolet theory, to avoid introducing dangerous sources of flavor violation. However, allowing for such an alignment opens up a a richer coupling structure compared to the vector case. 

Comparing the right and the left plot of Fig.~\ref{fig:gaee-constraints} one can see that the reach for the pseudo-scalar is substantially worse than for the scalar, to the extent that only phase II of \muthreee will be able to probe unconstrained parameter space. This is due to an unfortunate cancellation in the muon decay amplitude for CP-odd scalars emitted from the muon line. This cancellation can be easily understood by looking at the process at the amplitude level: in the $m_X\ll m_\mu$ limit, we can factorize the light particle emission from the muon line from the rest of the process by expanding in the virtuality of the off-shell muon. In this limit the CP-odd scalar emission goes to zero with respect to the scalar one because $\bar{u}(p-k) \gamma_5 u(p) \sim \mathcal{O}(k)$  whereas $\bar{u}(p-k)u(p) \sim \mathcal{O}(m_\mu)$. 

Currently, the most relevant existing bounds come from the E141 beam dump \cite{PhysRevLett.59.755} and the anomalous magnetic dipole moment of the muon $(g-2)_\mu$. The contribution to $(g-2)_\mu$ at one-loop are given by (see for example Refs.~\cite{Batell:2016ove,Freitas:2014pua})
\begin{align}
a_\mu^S &= \frac{g_S^{\mu\,2}}{8\pi^2}\int_0^1\diff x \frac{(1-x)^2(1+x)}{(1-x)^2+x\frac{m_X^2}{m_\mu^2}} \,,\\
a_\mu^P &= -\frac{g_P^{\mu\,2}}{8\pi^2}\int_0^1 \diff x \frac{x^3}{x^2+(1-x)\frac{m_X^2}{m_\mu^2}} \,,
\end{align} 
for the cases where $X$ is either a CP-even scalar or a CP-odd scalar. As these contributions come with opposite signs we will treat them separately. To date, there is still a substantial tension between the SM predictions and the measurement, as well as between different theoretical predictions. To characterize this discrepancy we show for the CP even scalar case two different results: \emph{i)} the best-fit region (at $\pm 3\sigma$) in green assuming the dispersive methods of extracting the hadronic vacuum polarization (HVP) contributions are correct~\cite{Aoyama:2020ynm}, \emph{ii)} the $2\sigma$ upper limit in black using the recent BMW result for HVP~\cite{Borsanyi:2020mff}. 
In formulas, we define 
\begin{align}
    \Delta a_\mu^\text{disp} &= a^\text{exp}_\mu - a^\text{disp}_\mu = 249(48) \times 10^{-11}\,,\label{eq:disp}\\
     \Delta a_\mu^\text{lattice} &= a^\text{exp}_\mu - a^\text{lattice}_\mu = 105(62)\times 10^{-11}\,.\label{eq:lattice}
\end{align}
with $a^\text{disp}_\mu = 116 591 810(43)\times 10^{-11}$ \cite{Aoyama:2020ynm}, $a^\text{lattice}_\mu = 116592160(57)\times 10^{-11}$ and the recent FNAL update $a^\text{exp}_\mu = 116 592 055(24)\times10^{-11}$ \cite{Muong-2:2023cdq} where the theory and experimental errors are added in quadrature. Due to the leading sign of the pseudoscalar case the contribution $a_\mu^P$ always increases the tension between experiment and dispersive predictions. Consequently, we show two conservative limits: \emph{i)} the $5\sigma$ limit in black where the maximal contribution from the axion equal to the size of the BMW lattice uncertainty and \emph{ii)} the future case in gray where the experimental uncertainty determines the allowed size of the pseudoscalar couplings again at $5\sigma$.

\begin{table}
    \centering
  \begin{tabular}{ C{2.5cm} C{2.5cm} C{1.cm}}
    \toprule
    			Scalar  					&Pseudo-scalar				& Figure  \\
			($g^{\mu,e}_P=0$)			& ($g^{\mu,e}_S=0$)				&   \\ \midrule
  $g_S^e=\num{5e-5}$	& $g_P^e=\num{5e-5}$ 		& Fig.~\ref{fig:gaee-constraints}		\\
 $g_S^e=m_e/m_\mu g_{S}^\mu$	& $g_P^e=m_e/m_\mu g_P^\mu$ 		&Fig.~\ref{fig:gaee-constraints_mass_hierarchy}	\\
   $g_S^\mu=0$	& $g_P^\mu=0$ 		&	Fig.~\ref{fig:gaee-constraints_gee_only}	\\\toprule
    \end{tabular}
\caption{List of benchmarks considered for the scalar and pseudoscalar model. Fig.~\ref{fig:gaee-constraints_mass_hierarchy} and Fig.~\ref{fig:gaee-constraints_gee_only} can be found in Appendix~\ref{app:additional_results}. \label{tab:benchmarks} }
\end{table}

Alternative scenarios for the scalar couplings are summarized in Table~\ref{tab:benchmarks} and discussed in more detail in Appendix~\ref{app:additional_results}. In a nutshell, Fig.~\ref{fig:gaee-constraints_mass_hierarchy} shows that a prompt search at \muthreee is not the best way to constrain the theoretically appealing scenarios where $g_{S,A}^e/g_{S,A}^\mu=m_e/m_\mu$. This is because the scalar $X$ is already too long-lived in the regime that is still allowed by existing bounds.  We leave an investigation of a displaced search and a comparison with the reach of other experiments \cite{Galon:2022xcl,Gninenko:2300189,Ilten:2015hya,Baltzell:2022rpd} for future work. Fig.~\ref{fig:gaee-constraints_gee_only} shows that when the muon coupling is negligible with respect to the electron coupling, the \muthreee expected reach is always surpassed by the present bounds on $\pi^+\to e^+\nu X$~\cite{Altmannshofer:2022ckw} with $X\to e^+e^-$, which benefits from the chirality suppression of the SM background $\pi^+\to e^+\nu\gamma^*$ with $\gamma^*\to e^+e^-$. 

\section{Conclusions and outlook}
In summary, \muthreee has a well-established capability to detect or constrain light $e^+e^-$ resonances, particularly dark photons. In this study, we systematically explored alternative models that produce this signature and found that the most compelling cases are associated with the dark photon and (pseudo)scalars primarily coupling to muons. We also investigated the impact of kinematic variables, such as track angles and energy spectra, on enhancing sensitivity in a standard bump-hunt search. The inclusion of two such variables demonstrated an approximately twofold improvement in sensitivity to the $\mu^+ \to X e^+ \bar\nu_\mu \nu_e$ branching ratio, without requiring detector upgrades or changes in data collection strategies.

Our proposed analysis may extend to the study of exotic decays involving other Standard Model particles, where a light particle is emitted from SM muons and decays into a dilepton pair. An example of this is the $K^+\to\mu^+\nu X$ decays mentioned in Ref.~\cite{Goudzovski:2022vbt}. The potential reach of NA62 and future kaon facilities for these final states warrants further investigation.

However, our study has two significant limitations. Firstly, generating a high-statistics Monte Carlo background sample for a study that simultaneously bins in more than two kinematic variables, in addition to the $e^+e^-$ invariant mass, proved challenging. Nonetheless, in practical experimental scenarios, this limitation can be addressed by extracting background estimates from sidebands in the $e^+e^-$ invariant mass spectrum. Machine learning techniques, such as the CWoLa prescription \cite{Collins:2019jip}, may offer effective solutions to this challenge, potentially surpassing our current estimates.

Finally, our analysis focused solely on prompt decays of $X$. For scalar models, especially, this limitation is significant (see Appendix~\ref{app:additional_results}). The primary sources of background in a displaced analysis are expected to arise from photon conversions within the detector material and tracks from unrelated Michel decays that cross randomly. The former can be mitigated with a veto on regions in which there is detector material, while the latter is suppressed to an extent by the timing capabilities of the detector. The primary limitation of such a search is likely to be signal acceptance, a topic we defer to future work.

\section*{Acknowledgements}
We are grateful to Yongsoo Jho for collaboration in early stages of this work as well as for useful discussions throughout. We also thank Ann-Kathrin Perrevoort for feedback on \muthreee as well as Jure Zupan, Dean Robinson, Stephania Gori, Jeff Dror, Bertrand Echenard, Zoltan Ligeti and Francesca Acanfora for useful discussions. The work of SK was supported by the Office of High Energy Physics of the U.S. Department of Energy under contract DE-AC02-05CH11231. KL was in part supported by the Berkeley Center for Theoretical Physics. TO is supported by the DOE Early Career Grant DESC0019225. Part of this work was performed at the Aspen Center for Physics, which is supported by National Science Foundation grant PHY-1607611.

\appendix
%
\section{Additional Models}\label{app:additional_results}
\begin{figure*}
    \centering
    \includegraphics[width=0.495\linewidth]{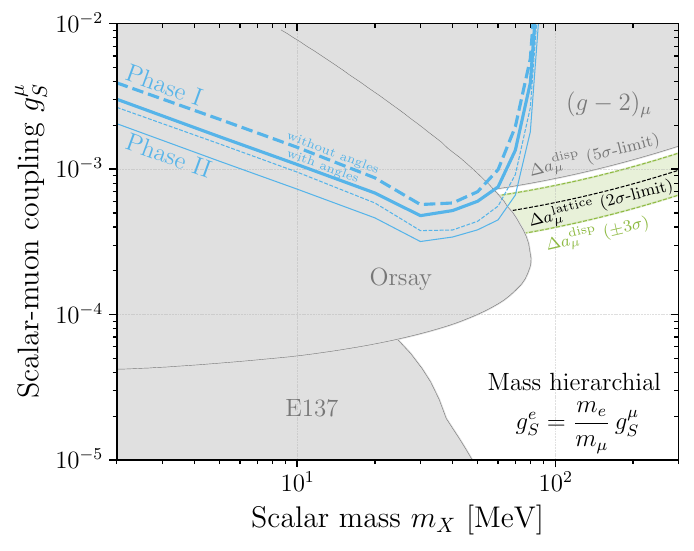}
    \includegraphics[width=0.495\linewidth]{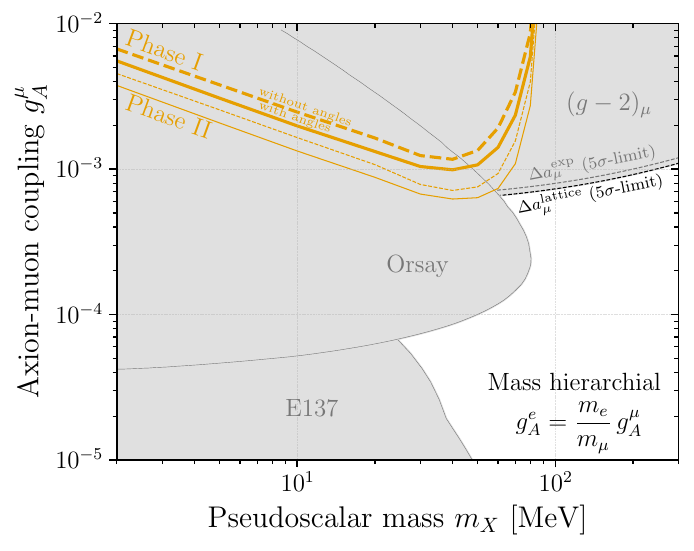}
    \caption{Parameter space of a  CP-even scalar ({\bf left}) and a CP-odd scalar ({\bf right}) with hierarchy between the muon and electron couplings equal to the SM mass hierarchy: $g_{S,A}^e/g^\mu_{S,A}=m_e/m_\mu$. The {\color{defgrey}\bf gray shaded} areas show constraints from beam dump experiments~\cite{Davier:1989wz,Bjorken:1988as} and the $(g-2)_\mu$ measurement. In the scalar case the best-fit region for $\Delta a_\mu^{\text{disp}}$ defined in Eq.~\eqref{eq:disp} is shown in {\color{defgreen}{\bf green}} while the limit from $\Delta a_\mu^{\text{lattice}}$ defined in Eq.~\eqref{eq:lattice} is indicated as a {\bf dashed black} line. Projections for the 95\% upper limit on $g_{S}^\mu$ {\color{cblblue}\bf light-blue} and  $g_{A}^\mu$ {\color{cborange}\bf orange} are weaker than current limits.}
    \label{fig:gaee-constraints_mass_hierarchy}
\end{figure*}

In this appendix, we provide additional information about potential models that could have been of interest for $\mu^+ \to X e^+ \bar\nu_\mu \nu_e$ but, as it turned out, were subjected to stronger constraints from other experiments. 

\begin{figure}
    \centering
        \includegraphics[width=1\linewidth]{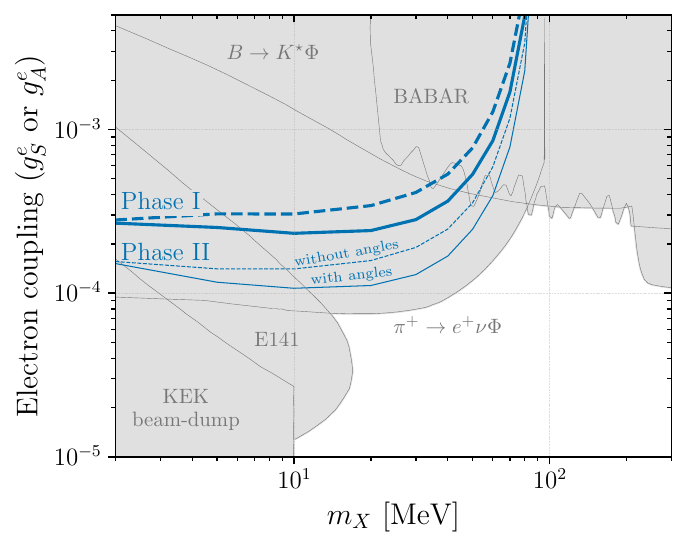}
    \caption{Parameter space of CP-even or a CP-odd scalar which couples exclusively to electrons/positrons. In this case the rates are exactly identical in the limit where the electron is massless. Projections for the 95\% upper limit at \muthreee shown in {\color{cbblue}\bf blue} are weaker than current constraints from pion decays derived in Ref.~\cite{Altmannshofer:2022ckw} from SINDRUM data~\cite{EICHLER1986101} ({\color{defgrey}\bf gray shaded}).}
    \label{fig:gaee-constraints_gee_only}
\end{figure}

First we further explore the parameter space of CP-even and CP-odd scalars considering the case $g_{S,P}^e/g^\mu_{S,P}=m_e/m_\mu$ in Fig.~\ref{fig:gaee-constraints_mass_hierarchy} and the one of pure electron coupling in Fig.~\ref{fig:gaee-constraints_gee_only}. In Fig.~\ref{fig:gaee-constraints_mass_hierarchy} the scalar particle tends to decay displaced leading to a reduction of the signal strength for $\mu^+ \to X e^+ \bar\nu_\mu \nu_e$ with a prompt $X\to e^+e^-$. The future \muthreee sensitivity is weaker than current beam dumps constraints from Orsay \cite{Davier:1989wz} and E137 \cite{Bjorken:1988as}. The suppression is exacerbated for CP-odd scalar because of their reduced rate compared to the CP-even scalar as explained in Sec.~\ref{sec:lightscalar}. In Fig.~\ref{fig:gaee-constraints_gee_only} the constraints from pion decays derived from SINDRUM data~\cite{EICHLER1986101} are always stronger than the \muthreee sensitivity because of the chirality suppression of the corresponding SM background~\cite{Altmannshofer:2022ckw}. 

\begin{figure}
    \centering
    \includegraphics[width=1\linewidth]{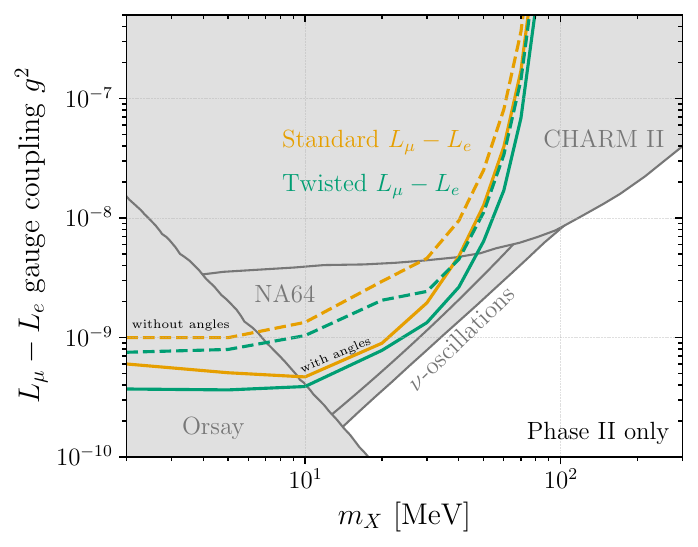}
    \caption{Parameter space of the standard and twisted $L_\mu - L_e$ gauge bosons defined in Eq.~\eqref{eq:LmuLenormal} and Eq.~\eqref{eq:LmuLetwisted}. Projections for the 95\% upper limit at \muthreee shown in {\color{cborange}\bf orange} and {\color{cbgreen}\bf green} respectively are weaker than current constraints from NA64~\cite{NA64:2023ehh} and neutrino scattering and oscillation experiments~\cite{TEXONO:2006xds,Borexino:2017fbd,Coloma:2020gfv} ({\color{defgrey}\bf gray shaded})}
    \label{fig:Lmu_minus_Le_limits}
\end{figure}

Second, we explore models of light vectors gauging $U(1)_{L_\mu-L_e}$ whose current can be  
\begin{equation}\label{eq:LmuLenormal}
    J_{L_\mu-L_e}^\alpha = (L_\mu^\dagger \bar\sigma^\alpha L_\mu - \bar{\mu}^\dagger \bar\sigma^\alpha \bar{\mu}  ) -  (L_e^\dagger \bar\sigma^\alpha L_e - \bar{e}^\dagger \bar\sigma^\alpha \bar{e}  )\, ,    
\end{equation}
or 
\begin{equation}
\tilde{J}_{L_\mu-L_e}^\alpha = (L_\mu^\dagger \bar\sigma^\alpha L_\mu + \bar{\mu}^\dagger \bar\sigma^\alpha \bar{\mu}  ) -  (L_e^\dagger \bar\sigma^\alpha L_e + \bar{e}^\dagger \bar\sigma^\alpha \bar{e})\,,\label{eq:LmuLetwisted}    
\end{equation}
the later of which we dub ``twisted'' $U(1)_{L_\mu-L_e}$ in what follows. These models provide an example of a new resonance which also couples to neutrinos hence generalizing the simplified lagrangian of Eq.~\eqref{eq:Lagrangians}. Fig.~\ref{fig:Lmu_minus_Le_limits} shows how the expected reach at \muthreee is weaker than the present constraints from neutrino scattering and oscillation experiments~\cite{Bilmis:2015lja,Lindner:2018kjo,Coloma:2020gfv} and NA64~\cite{NA64:2023ehh}. Note that the neutrino scattering experiments are slightly weaker than the oscillation constraints and are therefore omitted from the figure.

In general the neutrino constraints (see for example Ref.~\cite{Coloma:2022umy} for a summary) together with the existing constraints on the electron coupling itself tend to make the of neutrino coupling small enough to be neglected in the signal rate for $\mu^+ \to X e^+ \bar\nu_\mu \nu_e$ also for general scalar mediator. This observation justifies a posteriori the parametrization of Eq.~\eqref{eq:Lagrangians}. 

\section{Impact of muon polarization}
\label{app:Polarization}
\begin{figure*}
    \centering
    \includegraphics[width = \textwidth]{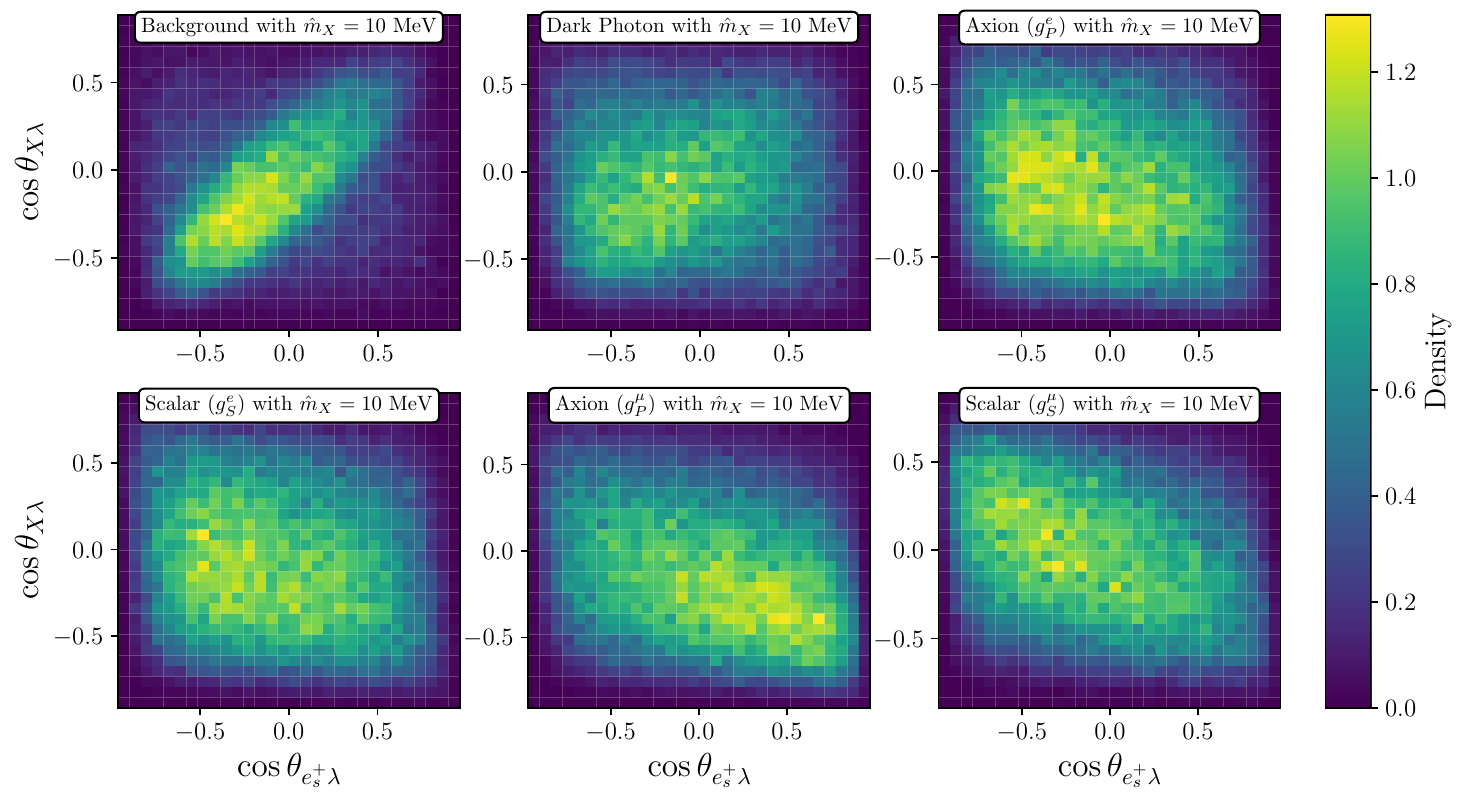}
    \caption{The distributions over the angles $\cos \theta_{e_s^+\lambda}$ and $\cos \theta_{X \lambda}$ for $\hat{m}_X = \SI{10}{\MeV}$  with $10^5$ simulated events for the five different signal models plus the SM background.}
    \label{fig:Polarization_Angles}
\end{figure*}

Accounting for the  muon polarization is potentially useful when using a muon beam obtained from pion decaying at rest, as is the case in the PSI beam line. The theoretical expectation is that these ``surface'' muons are  100\% polarized in the opposite direction of the momentum vector. In practice, the polarization of the stopped muons depends on depolarization effects and should be determined experimentally. As a reference for this study we use the measured averaged polarization at the MEG experiment~\cite{MEG:2015kvn} at PSI which is 86\% in the opposite direction of the momentum vector. The only detail which is new when accounting for polarization is the inclusion of two additional angles; the angle between the momentum of $e_s^+$ and the polarization vector $\lambda$, which we call $\cos \theta_{e_s^+ \lambda}$, and the angle between the momentum of $X$ and the polarization vector, which we call $\cos \theta_{X \lambda}$. The distributions with respect to these two variables show mild differences depending on the signal model as can be seen in Fig. \ref{fig:Polarization_Angles}. 

\begin{figure}
    \centering
    \includegraphics[width=0.5\textwidth]{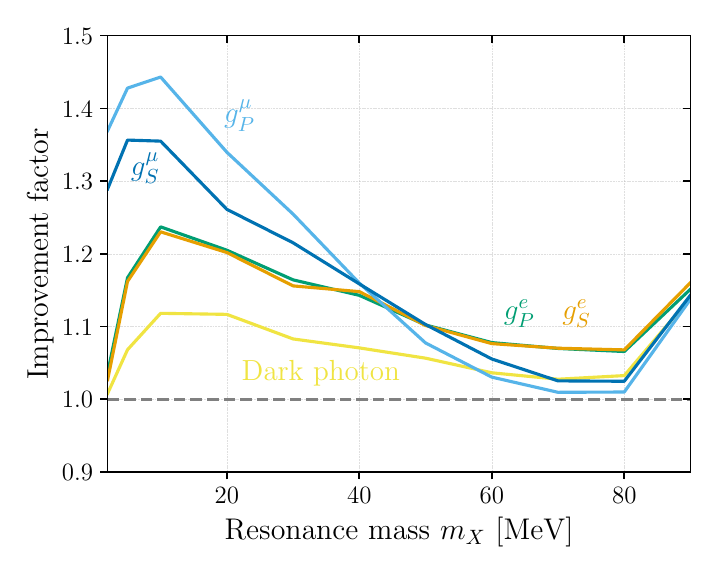}
    \caption{Improvement of the reach shown as a ratio of the rates (couplings squared) using the polarization dependent angles $\cos \theta_{e_s^+\lambda}$ and $\cos \theta_{X\lambda}$. The same models as Fig. \ref{fig:Improvement_Factor} are shown.}
    \label{fig:Polarization_Improvement}
\end{figure}

Although the dependence on the polarization is modest, it does allow one to improve the constraints, as shown in Fig.~\ref{fig:Polarization_Improvement}. Moreover, we expect that the correlations between those variables and the polarization dependent $\cos \theta_{e_s^+\lambda}$ and $\cos \theta_{X\lambda}$ to be fairly mild. In a full, data-driven experimental analysis we therefore recommend the inclusion of variables which are sensitive to the muon polarization, in addition to the kinematical variables discussed in Sec.~\ref{sec:angles}.

\section{Simulation framework and validation}
\label{app:validation}
The reconstructed final state consists of two positrons and an electron with four-momenta $p_{p_1}$, $p_{p_2}$ and $p_e$, respectively. There are therefore two possible combinations comprising of an electron-positron pair, whose invariant masses are denoted with $m_{ee_1}$ and  $m_{ee_2}$. In Sec.~\ref{sec:analysis-strategy} our search strategy comprised of first performing a bump hunt in the $m_{ee}$ distribution before examining the angular distributions of the positron and the reconstructed new physics state $X$ to further extract signal from background. 
\subsection{Truth-level simulation}

\begin{figure*}
\centering
\begin{minipage}{1.\textwidth}
\includegraphics[width = 0.495\textwidth]{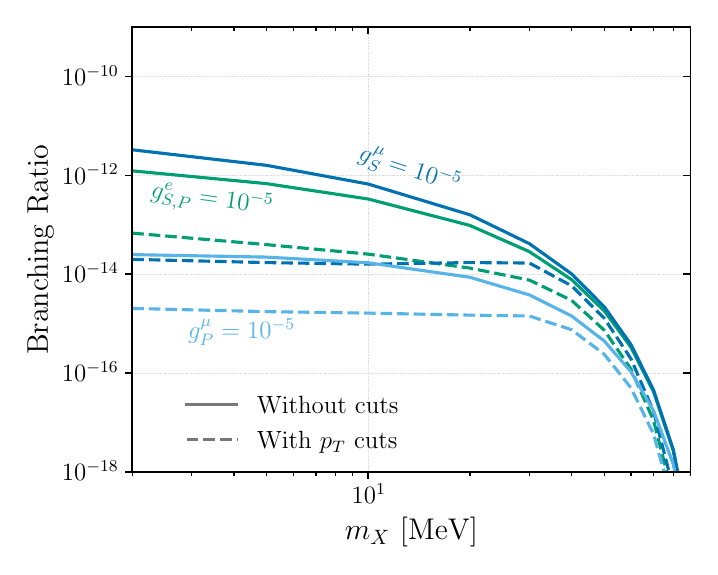}
\includegraphics[width = 0.495\textwidth]{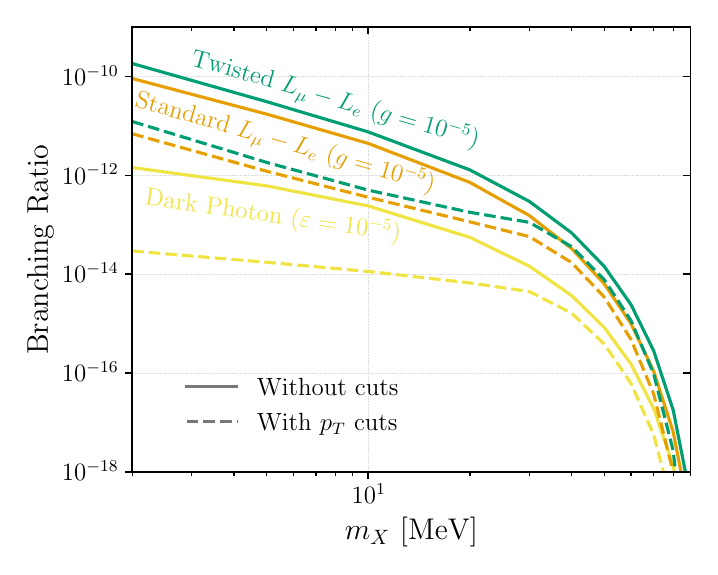}
\end{minipage}
\caption{The branching ratios for $\mu^+ \to X e^+ \bar\nu_\mu \nu_e,\, X\to e^+ e^-$, with and without $p_T>10$ MeV cut on all tracks, for the relevant coupling set to $g=10^{-5}$. \textbf{Left:} Spin 0 particles with a coupling predominantly to either electrons and muons. For the electron coupling, the scalar and pseudo-scalar cases have identical branching ratios, up to $m_e/m_\mu$ suppressed corrections. For muon coupling, the branching ratio to a pseudo-scalar is suppressed relative to the branching ratio to the scalar, as explained in Sec.~\ref{sec:results}. \textbf{Right:} Spin 1 particles, specifically the dark photon and both $L_\mu-L_e$ models.
 The different models are labeled the same as in Fig. \ref{fig:Improvement_Factor} and \ref{fig:Lmu_minus_Le_limits}.\label{fig:branching_ratios}} 
\label{fig:BRs}
\end{figure*}
\begin{figure*}
\centering
\begin{minipage}{1.\textwidth}
\includegraphics[width = 0.495\textwidth]{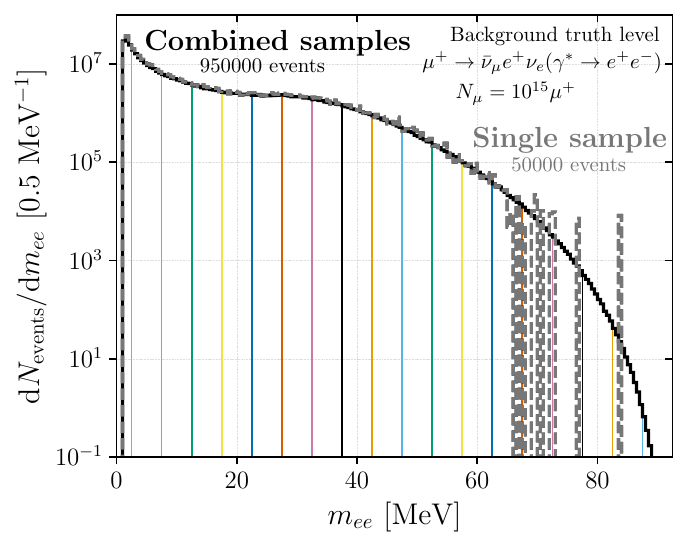}
\includegraphics[width = 0.495\textwidth]{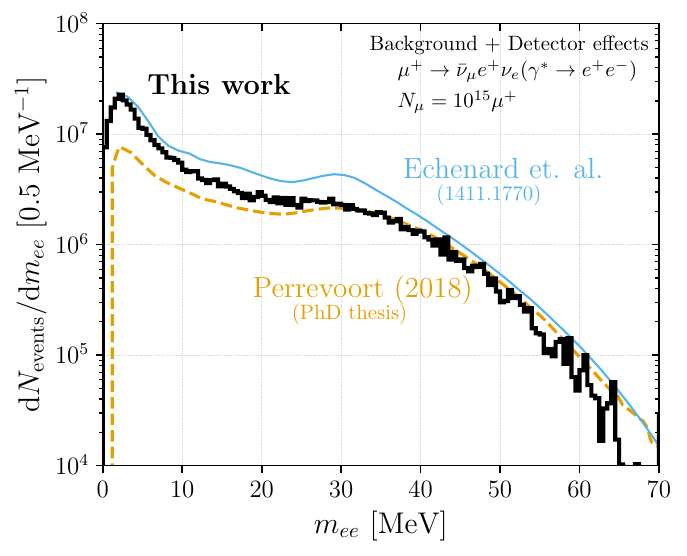}
\end{minipage}
\caption{Invariant mass distribution $m_{ee} \equiv m_{ee,1}+m_{ee,2}$ for the dominant Standard Model background, internal conversion $\mu^+ \to \bar{\nu}_\mu e^+ \nu_e (\gamma^* \to e^+e^-)$. (\textbf{Left:}) Validation of our phase-space slicing method to generate sufficient statistics across the full kinematic range of the invariant mass. Colored histograms are the individual samples centered around a mass hypothesis \mbox{$|m_{ee,i}-m_X|\leq\SI{2.5}{\MeV}$}, with $i=1,2$, while the black histogram is the sum over these samples. The gray histogram is from a single, inclusive background sample. \textbf{(Right:)} The full background $m_{ee}$-distribution (solid-black) including detector effects (see \cref{sec:detector-effects}). We also show the background distributions from Refs.~\cite{Perrevoort:2018okj} (thin-orange) and \cite{Echenard:2014lma} (thin-blue). For both panels we take the total number of muon decays $N_\mu = 10^{15}$.} 
\label{fig:bkg-validation}
\end{figure*}
To simulate signal events, we used the  \texttt{FeynRules 2.3} \cite{Alloul:2013bka} to generate a \texttt{Universal FeynRules Output} \cite{Degrande:2011ua} model file for the models defined in the previous section. We then generated events using \texttt{MadGraph5\_aMC@NLO 3.4.1} \cite{Alwall:2014hca}, taking care to specify the polarization of the muon. In addition we have also analytically computed the $\mu^+\to e^+ \bar\nu_\mu \nu_e X$ decay rates for the various signal models.
With these amplitudes and our own Monte-Carlo tool, we have validated the differential decay distributions obtained with MadGraph. The branching ratios for the various models are shown in Fig.~\ref{fig:branching_ratios}, with and without $p_T$ cuts on the tracks. 

The dominant background is the internal conversion process \mbox{$\mu^+\to e^+ e^- e^+ \bar\nu_\mu \nu_e$} \cite{Echenard:2014lma,Perrevoort:2018okj}, which we also simulate with MadGraph. (See Ref.~\cite{Flores-Tlalpa:2015vga} for the relevant amplitude and differential decay rates of this process.) In order to perform the search over the full range of kinematically accessible masses $m_X$, and study the subsequent kinematic variables in Sec.~\ref{sec:angles}, a sufficiently large number of events must be generated. This is largely through the difficulty in sampling the high-mass endpoint of the $m_{ee}$ distribution as this corresponds to an extreme configuration in five-body phase space where the neutrinos are very soft. It is therefore not feasible to generate a single, statistically satisfactory background sample. Instead we generate a separate background sample for each mass hypothesis of the dark sector particle ($m_X$). We do so by demanding that either \mbox{$|m_{ee,1}-m_X|\leq\SI{2.5}{\MeV}$} or \mbox{$|m_{ee,2}-m_X|\leq\SI{2.5}{\MeV}$} at the generator level, and weight the sample with its corresponding fiducial width. This cut is not available in the standard MadGraph cards and thus requires a modification to the MadGraph code responsible for imposing kinematic cuts. Our implementation can be found on \href{https://github.com/tobyopfer/mu3e-light-new-physics/tree/main}{GitHub} \href{https://github.com/tobyopfer/mu3e-light-new-physics/tree/main}{\textcolor{gray}{\faGithub}}, while in the left-hand side panel of Fig.~\ref{fig:bkg-validation} we validate this approach. Here we show in grey-dashed the $m_{ee}$ distribution (including both possible $e^+e^-$-pairs) for a single run of MadGraph generating 50000 events. At large invariant mass there is a clear degradation in the statistics due to the extreme configuration of phase-space that is required. This is overcome through the above phase-space slicing, shown as the colored histograms, producing a smooth distribution even towards the kinematic endpoint at large $m_{ee}$. Lastly we note that for all the background and signal samples we place a generator cut of $p_T \geq \SI{10}{\MeV}$ on all the electrons/positrons. 

Muons produced at the $\pi$E5 beam line at PSI are roughly 86\% polarized as measured by the MEG experiment \cite{MEG:2015kvn}. Since the muons at \muthreee are expected to be similarly polarized, we randomly mirror 7\% of the events in our samples along the polarization axis of the muon. We also assume the muons decay uniformly at rest within the target material of the experiment. 

\subsection{Detector modeling}
\label{sec:detector-effects}
\begin{figure}
\centering
\includegraphics[width = 0.5\textwidth]{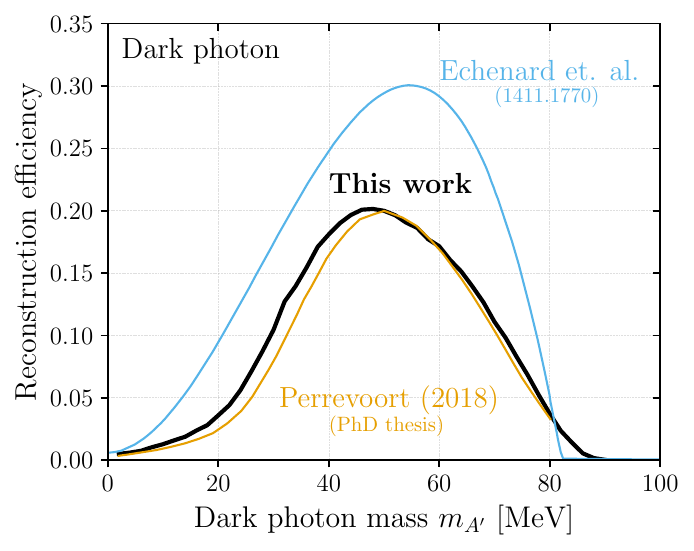}
\caption{Reconstruction efficiency of the signal events as a function of the dark photon mass. For comparison we also show the efficiency curves of Refs.~\cite{Perrevoort:2018okj} ({\color{cborange}\bf thin-orange}) and \cite{Echenard:2014lma} ({\color{cblblue}\bf thin-light-blue}).}
\label{fig:reco_eff}
\end{figure}
\begin{table*}
    \centering
  \begin{tabular}{l |  C{1.5cm} C{1.5cm} C{1.5cm} | C{1.5cm} C{1.5cm} | C{2cm} C{2cm} }
    \toprule
         \multirow{2}{*}{Cuts} & \multicolumn{3}{c |}{Phase space only} & \multicolumn{2}{c |}{Internal conv. bkg.} & \multicolumn{2}{c}{Dark photon  {\scriptsize($m_{A^\prime}=\SI{40}{\MeV}$)} } \\
         & Ref.~\cite{Perrevoort:2018okj} &  Ref.~\cite{Mu3e:2020gyw}  & This work & Ref.~\cite{Perrevoort:2018okj} & This work & Ref.~\cite{Perrevoort:2018okj} & This work \\\midrule
         $p_T \geq \SI{12}{\MeV}$ & $-$  & $-$ & $73.7\%$  & $-$ & $2.78\%$ & $-$ & $64.5\%$ \\
         & & & $(73.7\%)$ & & $(2.78\%)$ & & $(64.5\%)$ \\
         Geometric acceptance (short tracks) & $43.3\%$ & $38.1\%$ & $62.4\%$  & $-$ & $1.87\%$ & $-$ & $47.1\%$\\
         & & & ($84.7\%$)& & ($67.4\%$) & & $(73.1\%)$\\
         Rescaling factor  & $-$ & $-$ & $38.1\%$ & $-$ & $1.14\%$ & $-$ & $28.8\%$\\ 
         & & & $(61.1\%)$ & & $(61.1\%)$ & & $(61.1\%)$  \\ 
         Track reconstruction (short tracks) & $38.9\%$ & $34.1\%$ & $31.6\%$ & $0.51\%$ & $0.68\%$ & $17.5\%$ & $19.4\%$ \\ 
         & ($89.9\%$) & ($89.5\%$) & ($82.8\%$) & $(0.51\%)$ & $(59.1\%)$ & $(17.5\%)$ & $(67.5\%)$ \\
         Vertex quality and timing & $26.2\%$ & $\sim27.9\%^a$ & $29.4\%$ & $0.46\%$ & $0.63\%$ & $16.5\%$ & $18.1\%$  \\ 
          & $(93.1\%)$ & $(81.7\%)$ & $(93.1\%)$ & $(90.0\%)$ & $(93.1\%)$ & $(94.3\%)$ & $(93.1\%)$ \\ \midrule    
         Final efficiency & $36.2\%$ & $27.9\%$ & $29.4\%$ & $0.46\%$ & $0.63\%$ & $16.5\%$ & $18.1\%$ \\\toprule
         \multicolumn{7}{l}{\footnotesize ${}^a$Extracted assuming long-track efficiency for vertexing} \\
    \end{tabular}
    \caption{Cut-flow table for the various factors affecting the final detector efficiency. The percentages given are the running total efficiency while the numbers in parentheses are the percentage of events lost relative to the previous step. Note that here we choose to apply an additional rescaling factor to match the geometric acceptance, which is tuned to the flat phase space events and validated against both signal and background efficiencies. We use a hyphen ($-$) to denote steps that either are not applicable (the $p_T$ cut or rescaling factor for Refs.~\cite{Perrevoort:2018okj,Mu3e:2020gyw}) or where information is absent. See text for additional details and \cref{fig:reco_eff} for signal efficiency at alternative dark photon masses.}
    \label{tab:cut-flow}
\end{table*}

The \muthreee experiment consists of four layers of pixel trackers around a large conical target in which the muons are stopped, along with timing layers to assist in background rejection. It has sizeable angular coverage for electron/positron tracks with transverse momenta larger than approximately $\SI{10}{\MeV}$, while the neutrinos escape as missing energy. 

\muthreee separates the observed charged leptons into \textit{short tracks} and \textit{long tracks}. Roughly, the difference is that short tracks pass through four layers of pixel trackers, be that traversing the detector once, or for low $p_T$ tracks passing only through the two inner pixel layers before being bent back where they traverse back through the same layers again. For long tracks at least two more hits in the pixel layers are required. Effectively, these additional hits allow for a more precise measurement of their momentum vector as compared to that of the short tracks. The track reconstruction efficiencies for short track and long tracks have been simulated by the \muthreee collaboration, and are parametrized as a function of the total track momentum ($\mathbf{p}$) and its angle with the beam axis $(\theta)$ (Figs.~(19.1) and (19.4) in Ref.~\cite{Mu3e:2020gyw}). These parametrizations describe the probability that a track is reconstructed, \emph{if} it passed through at a tracking layer at least four times. It therefore does not include the geometric acceptance, which is defined as the probability of passing through the pixel layers at least four times. We must model this separately, as we describe below. 

To model the geometric acceptance we first begin by placing a $p_T$ cut of $\SI{12}{\MeV}$ on the electron/positron tracks, which rejects all charged leptons which would not make it through the first two pixel layers given the uniform $\SI{1}{\tesla}$ magnetic field in the detector. The remaining loss of tracks occurs as the length of the target in the beam-direction is similar to that of the inner pixel tracker layers. Consequently if a muon is stopped at the start/end of the target there is a sizeable probability that the lower $p_T$ tracks are sufficiently bent to miss the start/end of the inner pixel layer. For muons that decay more centrally this probability is effectively zero as all that matters is the gyro-radius of the track, which is accounted for by the $p_T$ cut. To account for these effects we firstly use the aforementioned track reconstruction map in $\mathbf{p}$ versus $\theta$, however, rather than use the $z$-axis which gives the probability of reconstructing a track, we use the non-zero region to build an approximate acceptance map of the detector. The resulting efficiency, quoted in the second line in \cref{tab:cut-flow}, is effectively the ``best case'' efficiency, for a muon decaying at the center of the detector. This is therefore an overestimate as it does not yet account for the extended size of the target.
To address this remaining discrepancy we apply an additional efficiency factor to match the result of the \muthreee collaboration. Fortunately \muthreee provides this factor for Monte Carlo events uniformly distributed in phase space, see Tab.~(22.1) of Ref.~\cite{Mu3e:2020gyw}. This ends up being an additional factor of $61.1\%$ in our reconstruction efficiency.\footnote{See also Tab.~(4.1) of Ref.~\cite{Perrevoort:2018okj} for a similar value.} Once we tuned our efficiency with a uniform phase space distribution, we validate it on both our signal and background events, see the additional rows of \cref{tab:cut-flow} which show good agreement with those stated in Ref.~\cite{Perrevoort:2018okj}.
After this step we can then apply the track reconstructions efficiencies mentioned above, through the digitization of Figs.~(19.1) and (19.4) from Ref.~\cite{Mu3e:2020gyw}. The latter of which governs the probability that a short track event can be `upgraded' to a long track, which plays an important role in smearing below. The final step involves a number of cuts selecting for quality of the vertex and timing in a given event. We cannot simulate these effects, but instead rely on statements in Ref.~\cite{Perrevoort:2018okj} where its is claimed that timing is $\sim 98\%$ efficient, while vertex cuts are $\sim 95\%$ efficient. A complete breakdown of these steps and comparisons to both Refs.~\cite{Mu3e:2020gyw,Perrevoort:2018okj} is given in \cref{tab:cut-flow}, where a hyphen is used to indicate steps where specific information is either not given by the experimental collaboration or is not relevant for them as it included automatically in their full detector simulation. Here we see that tuning our analysis pipeline on the flat phase-space events yields good agreement when applied to both signal and background events. This is especially apparent in \cref{fig:reco_eff} which shows the efficiency for the dark photon signal model over a range of dark photon masses. 
\begin{figure*}
\centering
\begin{minipage}{1.\textwidth}
\includegraphics[width = 0.495\textwidth]{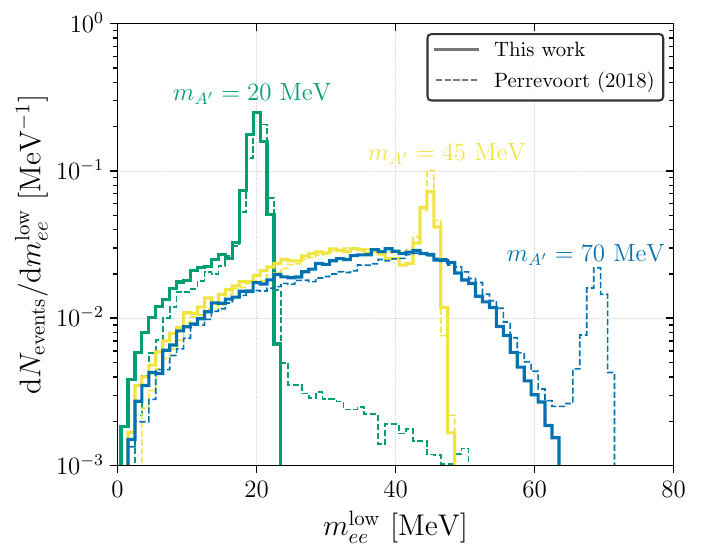}
\includegraphics[width = 0.495\textwidth]{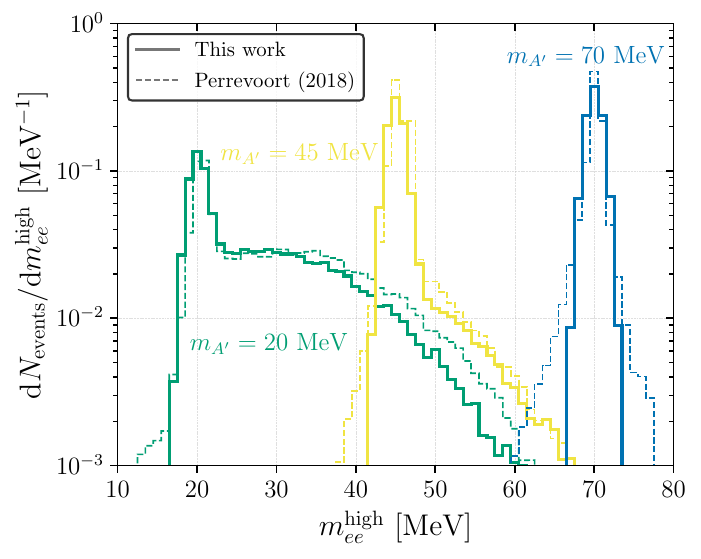}
\end{minipage}
\caption{Signal invariant mass distributions for the dark photon signal model. To compare to Ref.~\cite{Perrevoort:2018okj} (dashed lines) we show the low and high mass combinations of electron-positron pairs for $m_{A^\prime} = 20, 45, \SI{70}{\MeV}$ (green, yellow, blue).}
\label{fig:mee_smearing}
\end{figure*}

Importantly, the electron-positron pair invariant mass ($m_{ee}$) resolution depends not only on the momentum resolution but also on the angular resolution. In fact, it is this angular resolution which dominates the invariant mass resolution. Unfortunately, the \muthreee Technical Design Report (TDR) \cite{Mu3e:2020gyw} does not give the angular resolution of the tracks, but the invariant mass resolution for the positron-electron pair is reported in Ref.~\cite{Perrevoort:2018okj} as a function of the number of long tracks and the invariant mass value. We therefore use this result in our analysis, but since we are interested in angular correlations, our analysis would benefit from information about the angular resolution which we are unfortunately unable to incorporate. To determine the smearing in the $m_{ee}$ distribution we use Fig.~(6.12) in Ref.~\cite{Perrevoort:2018okj}, drawing from a Gaussian distribution with a width given by the reported RMS value which depends on the number of reconstructed long tracks. Note that we have implemented both $m_{ee}$ dependent and independent smearing (6.12a) versus (6.12b) of Ref.~\cite{Perrevoort:2018okj}, however, there is no appreciable difference between the two for the application at hand. To validate the smearing we show the distributions for both signal (\cref{fig:mee_smearing}) and background (right-hand panel of \cref{fig:bkg-validation}). For the background we see that there are sizeable differences at low-invariant masses. This discrepancy is larger than the mismatch in the efficiencies given in \cref{tab:cut-flow}, and likely points to additional effects not included in the rescaling factor to match the geometric acceptance. However, given that we over-predict background events in this region, our limits will simply be conservative at the bottom end of the mass range. For the signal smearing (\cref{fig:mee_smearing}) we see good agreement around the resonance peaks between the solid (our results) and dashed lines (Ref.~\cite{Perrevoort:2018okj}) for the different combinations of the electron-positron pairs invariant mass. However, away from the peaks we see that our distributions are narrower compared to Ref.~\cite{Perrevoort:2018okj}. This is because we assumed Gaussian smearing, when in reality a Crystal ball function with fatter tails would better describe the smearing, but it is not possible to implement given the information available. This also leads to the absence of a peak in the low-$m_{ee}$ combination for $m_{A^\prime} = \SI{70}{\MeV}$. Essentially a low-probability fluctuation is responsible for causing a signal event to be sufficiently smeared to appear in the low-invariant mass $m_{ee}$ distribution, which given our Gaussian modelling never occurs for our signal samples of $10^5$ events.   

\subsection{Additional Analysis details}
\label{app:analysis-details}

We are interested in performing a hypothesis test on each signal, parametrized by ($m_X,g$), with $g$ representing the coupling governing the signal strength for the model of interest ($\varepsilon,g^{\mu,e}_{S,P},g^{\mu,e}_{V,A}$) where the two relevant hypotheses are:
\begin{itemize}
    \item  $H_0$ = \{\textbf{Background only hypothesis:} Standard Model only\}
    \item  $H_1$ = \{\textbf{Signal hypothesis:} Signal model described by parameters $(m_X,g)$ exists\}
\end{itemize}
To test these hypotheses we define $\hat{m}_X$ as the invariant mass of the electron-positron pair closest to the hypothesis mass $m_X$ and take events in a window $m_X \pm \delta$ around this resonance. The width of this window, $\delta$, is chosen through maximization of the test statistic $S/\sqrt{B}$, which for our samples resulted in $\delta = \SI{1.5}{\MeV}$. The events that satisfy this cut are then binned in variables introduced in Sec.~\ref{sec:angles}. 

\begin{figure}
\centering
\includegraphics[width = 0.5\textwidth]{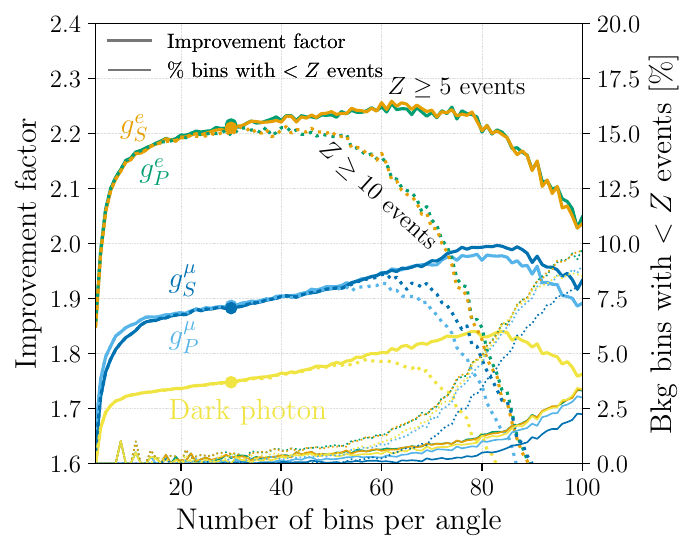}
\caption{Improvement factor including angular information in the reach projection as a function of the total number of bins in each angular direction for $m_X = \SI{40}{\MeV}$ and the angular observables $\cos\theta_{e_s^+ e_X^+}$ and $\cos\theta_{e^- e_s^+}$. The thick solid (dotted) lines indicate the improvement factor requiring at least 10 (5) Monte Carlo events with non-zero weights per bin for the background given non-zero signal events. The thin lines (with associated $y$-axis on the right-hand side) indicate the percentage of bins rejected (and therefore not contributing to the likelihood) given this requirement. The colors indicate the coupling structure of the signal model.}
\label{fig:improvement-vs-binning}
\end{figure}

As we are performing a binned likelihood analysis, the maximum number of angular bins we can consider faces a Monte Carlo limitation arising from our ability to sample these distributions across the entire domain of the angles while maintaining a satisfactory number of event per bin. Practically speaking the number of bins is therefore limited by the size of our Monte Carlo samples. This is primarily an issue for the background samples and is the key motivation for the splicing in phase-space. To demonstrate that our samples are sufficiently large, we show in \cref{fig:improvement-vs-binning} the improvement in reach from our angular analysis as a function of the number of bins per angle for several signal models. To ensure that bins with insufficient statistics are not driving the gain in sensitivity we discard bins from the likelihood that contain less than 10 (5) background Monte Carlo events with non-zero weights and at least one signal event; the corresponding increase in sensitivity is shown by the thick solid (thick dashed) curves. On the opposite $y$-axis we also show with thin lines the percentage of bins that are thrown away given this criteria. From the behavior we see that the gain in sensitivity quickly increases with a modest number of bins per angle before plateauing. Further improvement requires extremely fine binning with a relaxation of the criteria for the number of events with non-zero weights per bin. We therefore conclude our likelihood captures most of the relevant physics provided that we use $\sim$20 or more bins per variable. For the results presented in the main body of the paper a filled circle indicates our chosen binning (30 bins per angle), which we see is conservative compared to the finer binning results with fewer events per bin. For the (pseudo)scalar case coupled to electrons relaxing the event requirement per bin and increasing the number of bins from 30 to 60 results only in a $1.5\%$ improvement in the reach.    

\bibliographystyle{JHEP}
\bibliography{bibliography.bib}

\providecommand{\href}[2]{#2}\begingroup\raggedright\begin{thebibliography}{10}

\bibitem{Muong-2:2023cdq}
{\bf Muon g-2} Collaboration, D.~P. Aguillard et~al., {\it {Measurement of the
  Positive Muon Anomalous Magnetic Moment to 0.20 ppm}},
  \href{http://arxiv.org/abs/2308.06230}{{\tt arXiv:2308.06230}}.

\bibitem{Mu2e:2014fns}
{\bf Mu2e} Collaboration, L.~Bartoszek et~al., {\it {Mu2e Technical Design
  Report}},  \href{http://arxiv.org/abs/1501.05241}{{\tt arXiv:1501.05241}}.

\bibitem{COMET:2018auw}
{\bf COMET} Collaboration, R.~Abramishvili et~al., {\it {COMET Phase-I
  Technical Design Report}},  {\em PTEP} {\bf 2020} (2020), no.~3 033C01,
  [\href{http://arxiv.org/abs/1812.09018}{{\tt arXiv:1812.09018}}].

\bibitem{Teshima:2019orf}
N.~Teshima, {\it {Status of the DeeMe Experiment, an Experimental Search for
  $\mu$-$e$ Conversion at J-PARC MLF}},  {\em PoS} {\bf NuFact2019} (2020) 082,
  [\href{http://arxiv.org/abs/1911.07143}{{\tt arXiv:1911.07143}}].

\bibitem{MEGII:2018kmf}
{\bf MEG II} Collaboration, A.~M. Baldini et~al., {\it {The design of the MEG
  II experiment}},  {\em Eur. Phys. J. C} {\bf 78} (2018), no.~5 380,
  [\href{http://arxiv.org/abs/1801.04688}{{\tt arXiv:1801.04688}}].

\bibitem{Mu3e:2020gyw}
{\bf Mu3e} Collaboration, K.~Arndt et~al., {\it {Technical design of the phase
  I Mu3e experiment}},  {\em Nucl. Instrum. Meth. A} {\bf 1014} (2021) 165679,
  [\href{http://arxiv.org/abs/2009.11690}{{\tt arXiv:2009.11690}}].

\bibitem{Hesketh:2022wgw}
{\bf Mu3e} Collaboration, G.~Hesketh, S.~Hughes, A.-K. Perrevoort, and
  N.~Rompotis, {\it {The Mu3e Experiment}},  in {\em {2022 Snowmass Summer
  Study}}, 4, 2022.
\newblock \href{http://arxiv.org/abs/2204.00001}{{\tt arXiv:2204.00001}}.

\bibitem{Kuno:1999jp}
Y.~Kuno and Y.~Okada, {\it {Muon decay and physics beyond the standard model}},
   {\em Rev. Mod. Phys.} {\bf 73} (2001) 151--202,
  [\href{http://arxiv.org/abs/hep-ph/9909265}{{\tt hep-ph/9909265}}].

\bibitem{Okada:1999zk}
Y.~Okada, K.-i. Okumura, and Y.~Shimizu, {\it {Mu --\ensuremath{>} e gamma and
  mu --\ensuremath{>} 3 e processes with polarized muons and supersymmetric
  grand unified theories}},  {\em Phys. Rev. D} {\bf 61} (2000) 094001,
  [\href{http://arxiv.org/abs/hep-ph/9906446}{{\tt hep-ph/9906446}}].

\bibitem{Calibbi:2017uvl}
L.~Calibbi and G.~Signorelli, {\it {Charged Lepton Flavour Violation: An
  Experimental and Theoretical Introduction}},  {\em Riv. Nuovo Cim.} {\bf 41}
  (2018), no.~2 71--174, [\href{http://arxiv.org/abs/1709.00294}{{\tt
  arXiv:1709.00294}}].

\bibitem{Davidson:2020hkf}
S.~Davidson, {\it {Completeness and complementarity for $\mu \to e\gamma \mu
  \to e \bar e e$ and $\mu A \to eA$}},  {\em JHEP} {\bf 02} (2021) 172,
  [\href{http://arxiv.org/abs/2010.00317}{{\tt arXiv:2010.00317}}].

\bibitem{Bolton:2022lrg}
P.~D. Bolton and S.~T. Petcov, {\it {Measurements of $\mu\to 3e$ Decay with
  Polarised Muons as a Probe of New Physics}},
  \href{http://arxiv.org/abs/2204.03468}{{\tt arXiv:2204.03468}}.

\bibitem{Echenard:2014lma}
B.~Echenard, R.~Essig, and Y.-M. Zhong, {\it {Projections for Dark Photon
  Searches at Mu3e}},  {\em JHEP} {\bf 01} (2015) 113,
  [\href{http://arxiv.org/abs/1411.1770}{{\tt arXiv:1411.1770}}].

\bibitem{Calibbi:2020jvd}
L.~Calibbi, D.~Redigolo, R.~Ziegler, and J.~Zupan, {\it {Looking forward to
  lepton-flavor-violating ALPs}},  {\em JHEP} {\bf 09} (2021) 173,
  [\href{http://arxiv.org/abs/2006.04795}{{\tt arXiv:2006.04795}}].

\bibitem{Galon:2022xcl}
I.~Galon, D.~Shih, and I.~R. Wang, {\it {Dark photons and displaced vertices at
  the MUonE experiment}},  {\em Phys. Rev. D} {\bf 107} (2023), no.~9 095003,
  [\href{http://arxiv.org/abs/2202.08843}{{\tt arXiv:2202.08843}}].

\bibitem{Jho:2022snj}
Y.~Jho, S.~Knapen, and D.~Redigolo, {\it {Lepton-flavor violating axions at MEG
  II}},  {\em JHEP} {\bf 10} (2022) 029,
  [\href{http://arxiv.org/abs/2203.11222}{{\tt arXiv:2203.11222}}].

\bibitem{Hostert:2023gpk}
M.~Hostert, T.~Menzo, M.~Pospelov, and J.~Zupan, {\it {New physics in
  multi-electron muon decays}},  \href{http://arxiv.org/abs/2306.15631}{{\tt
  arXiv:2306.15631}}.

\bibitem{Hill:2023dym}
R.~J. Hill, R.~Plestid, and J.~Zupan, {\it {Searching for new physics at
  $\mu\rightarrow e$ facilities with $\mu^+$ and $\pi^+$ decays at rest}},
  \href{http://arxiv.org/abs/2310.00043}{{\tt arXiv:2310.00043}}.

\bibitem{Perrevoort:2018ttp}
{\bf Mu3e} Collaboration, A.-K. Perrevoort, {\it {The Rare and Forbidden:
  Testing Physics Beyond the Standard Model with Mu3e}},  {\em SciPost Phys.
  Proc.} {\bf 1} (2019) 052, [\href{http://arxiv.org/abs/1812.00741}{{\tt
  arXiv:1812.00741}}].

\bibitem{Bardin:1972qq}
D.~Y. Bardin, T.~G. Istatkov, and G.~Mitselmakher, {\it {On mu+
  ---\ensuremath{>} e+ nu(e) anti-nu(mu) e+ e- decay}},  {\em Yad. Fiz.} {\bf
  15} (1972) 284--287.

\bibitem{Flores-Tlalpa:2015vga}
A.~Flores-Tlalpa, G.~L\'opez~Castro, and P.~Roig, {\it {Five-body leptonic
  decays of muon and tau leptons}},  {\em JHEP} {\bf 04} (2016) 185,
  [\href{http://arxiv.org/abs/1508.01822}{{\tt arXiv:1508.01822}}].

\bibitem{Fael:2016yle}
M.~Fael and C.~Greub, {\it {Next-to-leading order prediction for the decay $
  \mu \to e\kern0.22em \left({e}^{+}{e}^{-}\right)\;\nu \kern0.2em
  \overline{\nu} $}},  {\em JHEP} {\bf 01} (2017) 084,
  [\href{http://arxiv.org/abs/1611.03726}{{\tt arXiv:1611.03726}}].

\bibitem{Pruna:2016spf}
G.~M. Pruna, A.~Signer, and Y.~Ulrich, {\it {Fully differential NLO predictions
  for the rare muon decay}},  {\em Phys. Lett. B} {\bf 765} (2017) 280--284,
  [\href{http://arxiv.org/abs/1611.03617}{{\tt arXiv:1611.03617}}].

\bibitem{Banerjee:2020rww}
P.~Banerjee, T.~Engel, A.~Signer, and Y.~Ulrich, {\it {QED at NNLO with
  McMule}},  {\em SciPost Phys.} {\bf 9} (2020) 027,
  [\href{http://arxiv.org/abs/2007.01654}{{\tt arXiv:2007.01654}}].

\bibitem{Fayet:1980ad}
P.~Fayet, {\it {Effects of the Spin 1 Partner of the Goldstino (Gravitino) on
  Neutral Current Phenomenology}},  {\em Phys. Lett. B} {\bf 95} (1980)
  285--289.

\bibitem{Okun:1982xi}
L.~B. Okun, {\it {LIMITS OF ELECTRODYNAMICS: PARAPHOTONS?}},  {\em Sov. Phys.
  JETP} {\bf 56} (1982) 502.

\bibitem{Georgi:1983sy}
H.~Georgi, P.~H. Ginsparg, and S.~L. Glashow, {\it {Photon Oscillations and the
  Cosmic Background Radiation}},  {\em Nature} {\bf 306} (1983) 765--766.

\bibitem{Holdom:1985ag}
B.~Holdom, {\it {Two U(1)'s and Epsilon Charge Shifts}},  {\em Phys. Lett. B}
  {\bf 166} (1986) 196--198.

\bibitem{Essig:2013lka}
R.~Essig et~al., {\it {Working Group Report: New Light Weakly Coupled
  Particles}},  in {\em {Community Summer Study 2013}: {Snowmass on the
  Mississippi}}, 10, 2013.
\newblock \href{http://arxiv.org/abs/1311.0029}{{\tt arXiv:1311.0029}}.

\bibitem{Proceedings:2012ulb}
{\em {Fundamental Physics at the Intensity Frontier}}, 5, 2012.

\bibitem{Jaeckel:2010ni}
J.~Jaeckel and A.~Ringwald, {\it {The Low-Energy Frontier of Particle
  Physics}},  {\em Ann. Rev. Nucl. Part. Sci.} {\bf 60} (2010) 405--437,
  [\href{http://arxiv.org/abs/1002.0329}{{\tt arXiv:1002.0329}}].

\bibitem{Pospelov:2008zw}
M.~Pospelov, {\it {Secluded U(1) below the weak scale}},  {\em Phys. Rev. D}
  {\bf 80} (2009) 095002, [\href{http://arxiv.org/abs/0811.1030}{{\tt
  arXiv:0811.1030}}].

\bibitem{Fayet:2007ua}
P.~Fayet, {\it {U-boson production in e+ e- annihilations, psi and Upsilon
  decays, and Light Dark Matter}},  {\em Phys. Rev. D} {\bf 75} (2007) 115017,
  [\href{http://arxiv.org/abs/hep-ph/0702176}{{\tt hep-ph/0702176}}].

\bibitem{Banerjee:2019hmi}
D.~Banerjee et~al., {\it {Improved limits on a hypothetical X(16.7) boson and a
  dark photon decaying into $e^+e^-$ pairs}},
  \href{http://arxiv.org/abs/1912.11389}{{\tt arXiv:1912.11389}}.

\bibitem{Perrevoort:2018okj}
A.-K. Perrevoort, {\em {Sensitivity Studies on New Physics in the Mu3e
  Experiment and Development of Firmware for the Front-End of the Mu3e Pixel
  Detector}}.
\newblock PhD thesis, U. Heidelberg (main), 2018.

\bibitem{Petersen:2023hgm}
{\bf FASER} Collaboration, B.~Petersen, {\it {First Physics Results from the
  FASER Experiment}},  in {\em {57th Rencontres de Moriond on Electroweak
  Interactions and Unified Theories}}, 5, 2023.
\newblock \href{http://arxiv.org/abs/2305.08665}{{\tt arXiv:2305.08665}}.

\bibitem{Bodas:2021fsy}
A.~Bodas, R.~Coy, and S.~J.~D. King, {\it {Solving the electron and muon $g-2$
  anomalies in $Z'$ models}},  \href{http://arxiv.org/abs/2102.07781}{{\tt
  arXiv:2102.07781}}.

\bibitem{Batley:2015lha}
{\bf NA48/2} Collaboration, J.~R. Batley et~al., {\it {Search for the dark
  photon in $\pi^0$ decays}},  {\em Phys. Lett.} {\bf B746} (2015) 178--185,
  [\href{http://arxiv.org/abs/1504.00607}{{\tt arXiv:1504.00607}}].

\bibitem{Lees:2014xha}
{\bf BaBar} Collaboration, J.~P. Lees et~al., {\it {Search for a Dark Photon in
  $e^+e^-$ Collisions at BaBar}},  {\em Phys. Rev. Lett.} {\bf 113} (2014),
  no.~20 201801, [\href{http://arxiv.org/abs/1406.2980}{{\tt
  arXiv:1406.2980}}].

\bibitem{Belle-II:2018jsg}
{\bf Belle-II} Collaboration, W.~Altmannshofer et~al., {\it {The Belle II
  Physics Book}},  {\em PTEP} {\bf 2019} (2019), no.~12 123C01,
  [\href{http://arxiv.org/abs/1808.10567}{{\tt arXiv:1808.10567}}]. [Erratum:
  PTEP 2020, 029201 (2020)].

\bibitem{Doria:2018sfx}
L.~Doria, P.~Achenbach, M.~Christmann, A.~Denig, P.~G\"ulker, and H.~Merkel,
  {\it {Search for light dark matter with the MESA accelerator}},  in {\em
  {13th Conference on the Intersections of Particle and Nuclear Physics}}, 9,
  2018.
\newblock \href{http://arxiv.org/abs/1809.07168}{{\tt arXiv:1809.07168}}.

\bibitem{Berlin:2018pwi}
A.~Berlin, S.~Gori, P.~Schuster, and N.~Toro, {\it {Dark Sectors at the
  Fermilab SeaQuest Experiment}},  {\em Phys. Rev. D} {\bf 98} (2018), no.~3
  035011, [\href{http://arxiv.org/abs/1804.00661}{{\tt arXiv:1804.00661}}].

\bibitem{Ilten:2018crw}
P.~Ilten, Y.~Soreq, M.~Williams, and W.~Xue, {\it {Serendipity in dark photon
  searches}},  {\em JHEP} {\bf 06} (2018) 004,
  [\href{http://arxiv.org/abs/1801.04847}{{\tt arXiv:1801.04847}}].

\bibitem{Krasznahorkay:2015iga}
A.~J. Krasznahorkay et~al., {\it {Observation of Anomalous Internal Pair
  Creation in Be8 : A Possible Indication of a Light, Neutral Boson}},  {\em
  Phys. Rev. Lett.} {\bf 116} (2016), no.~4 042501,
  [\href{http://arxiv.org/abs/1504.01527}{{\tt arXiv:1504.01527}}].

\bibitem{Nardi:2018cxi}
E.~Nardi, C.~D.~R. Carvajal, A.~Ghoshal, D.~Meloni, and M.~Raggi, {\it
  {Resonant production of dark photons in positron beam dump experiments}},
  {\em Phys. Rev. D} {\bf 97} (2018), no.~9 095004,
  [\href{http://arxiv.org/abs/1802.04756}{{\tt arXiv:1802.04756}}].

\bibitem{megIIatomki}
H.~Benmansour, ``{Search for the X17 particle with the MEG-II apparatus}.''
  \url{https://agenda.infn.it/event/28874/contributions/169428/attachments/93998/128533/ICHEP2022_Benmansour_0707.pdf},
  2022.
\newblock Online; accessed 20 November 2023.

\bibitem{Ilten:2015hya}
P.~Ilten, J.~Thaler, M.~Williams, and W.~Xue, {\it {Dark photons from charm
  mesons at LHCb}},  {\em Phys. Rev. D} {\bf 92} (2015), no.~11 115017,
  [\href{http://arxiv.org/abs/1509.06765}{{\tt arXiv:1509.06765}}].

\bibitem{Batell:2017kty}
B.~Batell, A.~Freitas, A.~Ismail, and D.~Mckeen, {\it {Flavor-specific scalar
  mediators}},  {\em Phys. Rev. D} {\bf 98} (2018), no.~5 055026,
  [\href{http://arxiv.org/abs/1712.10022}{{\tt arXiv:1712.10022}}].

\bibitem{Ballett:2018ynz}
P.~Ballett, S.~Pascoli, and M.~Ross-Lonergan, {\it {U(1)' mediated decays of
  heavy sterile neutrinos in MiniBooNE}},  {\em Phys. Rev. D} {\bf 99} (2019)
  071701, [\href{http://arxiv.org/abs/1808.02915}{{\tt arXiv:1808.02915}}].

\bibitem{DIMOPOULOS1979435}
S.~Dimopoulos, {\it A solution of the strong cp problem in models with
  scalars},  {\em Physics Letters B} {\bf 84} (1979), no.~4 435--439.

\bibitem{Holdom:1982ex}
B.~Holdom and M.~E. Peskin, {\it {Raising the Axion Mass}},  {\em Nucl. Phys.
  B} {\bf 208} (1982) 397--412.

\bibitem{Flynn:1987rs}
J.~M. Flynn and L.~Randall, {\it {A Computation of the Small Instanton
  Contribution to the Axion Potential}},  {\em Nucl. Phys. B} {\bf 293} (1987)
  731--739.

\bibitem{Rubakov:1997vp}
V.~Rubakov, {\it {Grand unification and heavy axion}},  {\em JETP Lett.} {\bf
  65} (1997) 621--624, [\href{http://arxiv.org/abs/hep-ph/9703409}{{\tt
  hep-ph/9703409}}].

\bibitem{Choi:1998ep}
K.~Choi and H.~D. Kim, {\it {Small instanton contribution to the axion
  potential in supersymmetric models}},  {\em Phys. Rev. D} {\bf 59} (1999)
  072001, [\href{http://arxiv.org/abs/hep-ph/9809286}{{\tt hep-ph/9809286}}].

\bibitem{Kilic:2009mi}
C.~Kilic, T.~Okui, and R.~Sundrum, {\it {Vectorlike Confinement at the LHC}},
  {\em JHEP} {\bf 02} (2010) 018, [\href{http://arxiv.org/abs/0906.0577}{{\tt
  arXiv:0906.0577}}].

\bibitem{Ferretti:2013kya}
G.~Ferretti and D.~Karateev, {\it {Fermionic UV completions of Composite Higgs
  models}},  {\em JHEP} {\bf 03} (2014) 077,
  [\href{http://arxiv.org/abs/1312.5330}{{\tt arXiv:1312.5330}}].

\bibitem{Belyaev:2015hgo}
A.~Belyaev, G.~Cacciapaglia, H.~Cai, T.~Flacke, A.~Parolini, and H.~Ser\^odio,
  {\it {Singlets in composite Higgs models in light of the LHC 750 GeV diphoton
  excess}},  {\em Phys. Rev. D} {\bf 94} (2016), no.~1 015004,
  [\href{http://arxiv.org/abs/1512.07242}{{\tt arXiv:1512.07242}}].

\bibitem{Bellazzini:2017neg}
B.~Bellazzini, A.~Mariotti, D.~Redigolo, F.~Sala, and J.~Serra, {\it {$R$-axion
  at colliders}},  {\em Phys. Rev. Lett.} {\bf 119} (2017), no.~14 141804,
  [\href{http://arxiv.org/abs/1702.02152}{{\tt arXiv:1702.02152}}].

\bibitem{Ferretti:2016upr}
G.~Ferretti, {\it {Gauge theories of Partial Compositeness: Scenarios for
  Run-II of the LHC}},  {\em JHEP} {\bf 06} (2016) 107,
  [\href{http://arxiv.org/abs/1604.06467}{{\tt arXiv:1604.06467}}].

\bibitem{Jeong:2018ucz}
K.~S. Jeong, T.~H. Jung, and C.~S. Shin, {\it {Axionic Electroweak
  Baryogenesis}},  {\em Phys. Lett. B} {\bf 790} (2019) 326--331,
  [\href{http://arxiv.org/abs/1806.02591}{{\tt arXiv:1806.02591}}].

\bibitem{Riordan:1987aw}
E.~M. Riordan et~al., {\it {A Search for Short Lived Axions in an Electron Beam
  Dump Experiment}},  {\em Phys. Rev. Lett.} {\bf 59} (1987) 755.

\bibitem{Davier:1989wz}
M.~Davier and H.~Nguyen~Ngoc, {\it {An Unambiguous Search for a Light Higgs
  Boson}},  {\em Phys. Lett. B} {\bf 229} (1989) 150--155.

\bibitem{Bjorken:1988as}
J.~D. Bjorken, S.~Ecklund, W.~R. Nelson, A.~Abashian, C.~Church, B.~Lu, L.~W.
  Mo, T.~A. Nunamaker, and P.~Rassmann, {\it {Search for Neutral Metastable
  Penetrating Particles Produced in the SLAC Beam Dump}},  {\em Phys. Rev. D}
  {\bf 38} (1988) 3375.

\bibitem{PhysRevLett.59.755}
E.~M. Riordan, M.~W. Krasny, K.~Lang, P.~de~Barbaro, A.~Bodek, S.~Dasu,
  N.~Varelas, X.~Wang, R.~Arnold, D.~Benton, P.~Bosted, L.~Clogher, A.~Lung,
  S.~Rock, Z.~Szalata, B.~W. Filippone, R.~C. Walker, J.~D. Bjorken,
  M.~Crisler, A.~Para, J.~Lambert, J.~Button-Shafer, B.~Debebe, M.~Frodyma,
  R.~S. Hicks, G.~A. Peterson, and R.~Gearhart, {\it Search for short-lived
  axions in an electron-beam-dump experiment},  {\em Phys. Rev. Lett.} {\bf 59}
  (Aug, 1987) 755--758.

\bibitem{Batell:2016ove}
B.~Batell, N.~Lange, D.~McKeen, M.~Pospelov, and A.~Ritz, {\it {Muon anomalous
  magnetic moment through the leptonic Higgs portal}},  {\em Phys. Rev. D} {\bf
  95} (2017), no.~7 075003, [\href{http://arxiv.org/abs/1606.04943}{{\tt
  arXiv:1606.04943}}].

\bibitem{Freitas:2014pua}
A.~Freitas, J.~Lykken, S.~Kell, and S.~Westhoff, {\it {Testing the Muon g-2
  Anomaly at the LHC}},  {\em JHEP} {\bf 05} (2014) 145,
  [\href{http://arxiv.org/abs/1402.7065}{{\tt arXiv:1402.7065}}]. [Erratum:
  JHEP 09, 155 (2014)].

\bibitem{Aoyama:2020ynm}
T.~Aoyama et~al., {\it {The anomalous magnetic moment of the muon in the
  Standard Model}},  {\em Phys. Rept.} {\bf 887} (2020) 1--166,
  [\href{http://arxiv.org/abs/2006.04822}{{\tt arXiv:2006.04822}}].

\bibitem{Borsanyi:2020mff}
S.~Borsanyi et~al., {\it {Leading hadronic contribution to the muon magnetic
  moment from lattice QCD}},  {\em Nature} {\bf 593} (2021), no.~7857 51--55,
  [\href{http://arxiv.org/abs/2002.12347}{{\tt arXiv:2002.12347}}].

\bibitem{Gninenko:2300189}
S.~Gninenko, {\it {Addendum to the NA64 Proposal: Search for the $A'\to
  invisible$ and $X\to e^+e^-$ decays in 2021}},  tech. rep., CERN, Geneva,
  2018.

\bibitem{Baltzell:2022rpd}
N.~Baltzell et~al., {\it {The Heavy Photon Search Experiment}},
  \href{http://arxiv.org/abs/2203.08324}{{\tt arXiv:2203.08324}}.

\bibitem{Altmannshofer:2022ckw}
W.~Altmannshofer, J.~A. Dror, and S.~Gori, {\it {New Opportunities for
  Detecting Axion-Lepton Interactions}},  {\em Phys. Rev. Lett.} {\bf 130}
  (2023), no.~24 241801, [\href{http://arxiv.org/abs/2209.00665}{{\tt
  arXiv:2209.00665}}].

\bibitem{Goudzovski:2022vbt}
E.~Goudzovski et~al., {\it {New physics searches at kaon and hyperon
  factories}},  {\em Rept. Prog. Phys.} {\bf 86} (2023), no.~1 016201,
  [\href{http://arxiv.org/abs/2201.07805}{{\tt arXiv:2201.07805}}].

\bibitem{Collins:2019jip}
J.~H. Collins, K.~Howe, and B.~Nachman, {\it {Extending the search for new
  resonances with machine learning}},  {\em Phys. Rev. D} {\bf 99} (2019),
  no.~1 014038, [\href{http://arxiv.org/abs/1902.02634}{{\tt
  arXiv:1902.02634}}].

\bibitem{EICHLER1986101}
R.~Eichler, L.~Felawka, N.~Kraus, C.~Niebuhr, H.~Walter, S.~Egli, R.~Engfer,
  C.~Grab, E.~Hermes, H.~Pruys, A.~{van der Schaaf}, D.~Vermeulen, W.~Bertl,
  N.~Lordong, U.~Bellgardt, G.~Otter, T.~Kozlowski, and J.~Martino, {\it Limits
  for short-lived neutral particles emitted in $\mu^+$ or $\pi^+$ decay},  {\em
  Physics Letters B} {\bf 175} (1986), no.~1 101--104.

\bibitem{Bilmis:2015lja}
S.~Bilmis, I.~Turan, T.~M. Aliev, M.~Deniz, L.~Singh, and H.~T. Wong, {\it
  {Constraints on Dark Photon from Neutrino-Electron Scattering Experiments}},
  {\em Phys. Rev. D} {\bf 92} (2015), no.~3 033009,
  [\href{http://arxiv.org/abs/1502.07763}{{\tt arXiv:1502.07763}}].

\bibitem{Lindner:2018kjo}
M.~Lindner, F.~S. Queiroz, W.~Rodejohann, and X.-J. Xu, {\it {Neutrino-electron
  scattering: general constraints on $Z'$ and dark photon models}},  {\em JHEP}
  {\bf 05} (2018) 098, [\href{http://arxiv.org/abs/1803.00060}{{\tt
  arXiv:1803.00060}}].

\bibitem{Coloma:2020gfv}
P.~Coloma, M.~C. Gonzalez-Garcia, and M.~Maltoni, {\it {Neutrino oscillation
  constraints on U(1)' models: from non-standard interactions to long-range
  forces}},  {\em JHEP} {\bf 01} (2021) 114,
  [\href{http://arxiv.org/abs/2009.14220}{{\tt arXiv:2009.14220}}]. [Erratum:
  JHEP 11, 115 (2022)].

\bibitem{NA64:2023ehh}
{\bf NA64} Collaboration, Y.~M. Andreev et~al., {\it {Probing Light Dark Matter
  with positron beams at NA64}},  \href{http://arxiv.org/abs/2308.15612}{{\tt
  arXiv:2308.15612}}.

\bibitem{Coloma:2022umy}
P.~Coloma, P.~Coloma, M.~C. Gonzalez-Garcia, M.~C. Gonzalez-Garcia, M.~Maltoni,
  M.~Maltoni, J.~a.~P. Pinheiro, J.~a.~P. Pinheiro, S.~Urrea, and S.~Urrea,
  {\it {Constraining new physics with Borexino Phase-II spectral data}},  {\em
  JHEP} {\bf 07} (2022) 138, [\href{http://arxiv.org/abs/2204.03011}{{\tt
  arXiv:2204.03011}}]. [Erratum: JHEP 11, 138 (2022)].

\bibitem{TEXONO:2006xds}
{\bf TEXONO} Collaboration, H.~T. Wong et~al., {\it {A Search of Neutrino
  Magnetic Moments with a High-Purity Germanium Detector at the Kuo-Sheng
  Nuclear Power Station}},  {\em Phys. Rev. D} {\bf 75} (2007) 012001,
  [\href{http://arxiv.org/abs/hep-ex/0605006}{{\tt hep-ex/0605006}}].

\bibitem{Borexino:2017fbd}
{\bf Borexino} Collaboration, M.~Agostini et~al., {\it {Limiting neutrino
  magnetic moments with Borexino Phase-II solar neutrino data}},  {\em Phys.
  Rev. D} {\bf 96} (2017), no.~9 091103,
  [\href{http://arxiv.org/abs/1707.09355}{{\tt arXiv:1707.09355}}].

\bibitem{MEG:2015kvn}
{\bf MEG} Collaboration, A.~M. Baldini et~al., {\it {Muon polarization in the
  MEG experiment: predictions and measurements}},  {\em Eur. Phys. J. C} {\bf
  76} (2016), no.~4 223, [\href{http://arxiv.org/abs/1510.04743}{{\tt
  arXiv:1510.04743}}].

\bibitem{Alloul:2013bka}
A.~Alloul, N.~D. Christensen, C.~Degrande, C.~Duhr, and B.~Fuks, {\it
  {FeynRules 2.0 - A complete toolbox for tree-level phenomenology}},  {\em
  Comput. Phys. Commun.} {\bf 185} (2014) 2250--2300,
  [\href{http://arxiv.org/abs/1310.1921}{{\tt arXiv:1310.1921}}].

\bibitem{Degrande:2011ua}
C.~Degrande, C.~Duhr, B.~Fuks, D.~Grellscheid, O.~Mattelaer, and T.~Reiter,
  {\it {UFO - The Universal FeynRules Output}},  {\em Comput. Phys. Commun.}
  {\bf 183} (2012) 1201--1214, [\href{http://arxiv.org/abs/1108.2040}{{\tt
  arXiv:1108.2040}}].

\bibitem{Alwall:2014hca}
J.~Alwall, R.~Frederix, S.~Frixione, V.~Hirschi, F.~Maltoni, O.~Mattelaer,
  H.~S. Shao, T.~Stelzer, P.~Torrielli, and M.~Zaro, {\it {The automated
  computation of tree-level and next-to-leading order differential cross
  sections, and their matching to parton shower simulations}},  {\em JHEP} {\bf
  07} (2014) 079, [\href{http://arxiv.org/abs/1405.0301}{{\tt
  arXiv:1405.0301}}].

\end{thebibliography}\endgroup

\end{document}